\documentclass[lettersize,journal]{IEEEtran}

\usepackage{stfloats}
\usepackage{verbatim}

\usepackage{cite}
\usepackage{amsmath,amssymb,amsfonts}
\usepackage{algorithmic}
\usepackage{graphicx}
\usepackage{textcomp}
\usepackage{xcolor}
\def\BibTeX{{\rm B\kern-.05em{\sc i\kern-.025em b}\kern-.08em
    T\kern-.1667em\lower.7ex\hbox{E}\kern-.125emX}}

\usepackage{balance}  
\usepackage{enumitem}

\usepackage{amsthm}
\usepackage{amssymb}
\usepackage{algorithm}
\usepackage{multirow}
\usepackage[hyphens]{url} 
\usepackage{caption}
\usepackage{url}
\usepackage{color}
\usepackage{cite}
\usepackage{times}
\usepackage{amsmath}
\usepackage{booktabs}

\newtheorem{theorem}{Theorem}

\usepackage{changepage}

\usepackage{lipsum}

\newlist{mylistenv}{enumerate}{3}
\newenvironment{mylist}[1]{%
    \setlist[mylistenv]{label=#1\arabic{mylistenvi}.,ref=#1\arabic{mylistenvi}}%
    \setlist[mylistenv,2]{label=#1\arabic{mylistenvi}.\arabic{mylistenvii}.,ref=#1\arabic{mylistenvi}.\arabic{mylistenvii}}%
    \setlist[mylistenv,3]{label=#1\arabic{mylistenvi}.\arabic{mylistenvii}.\arabic{mylistenviii}.,ref=#1\arabic{mylistenvi}.\arabic{mylistenvii}.\arabic{mylistenviii}}%
    \renewenvironment{mylist}{\begin{mylistenv}}{\end{mylistenv}}
    \begin{mylistenv}%
}{%
    \end{mylistenv}%
}

\usepackage{hyperref}
\usepackage{tikz}
\usepackage{pgfplots}
\pgfplotsset{compat=1.18}
\usepackage{diagbox}
\usepackage{cite}
\usetikzlibrary{positioning}
\usetikzlibrary{matrix}
\usetikzlibrary{patterns}
\usetikzlibrary{pgfplots.statistics, pgfplots.colorbrewer}
\usetikzlibrary{tikzmark}
\usepgfplotslibrary{colorbrewer}
\usepackage{pgfplotstable}

\usepackage{subfig}
\usepackage{graphicx,import}

\usepackage{array}
\newcolumntype{L}[1]{>{\raggedright\let\newline\\\arraybackslash\hspace{0pt}}m{#1}}
\newcolumntype{C}[1]{>{\centering\let\newline  \\\arraybackslash\hspace{0pt}}m{#1}}
\newcolumntype{R}[1]{>{\raggedleft\let\newline \\\arraybackslash\hspace{0pt}}m{#1}}

 \usepackage{mathptmx}      
\newif\ifICDE
\ICDEtrue
\ifICDE
\newcommand{\revised}[1]{\textcolor{black}{#1}}
\else
\newcommand{\revised}[1]{\textcolor{black}{#1}}
\fi

\newif\ifTKDE
\TKDEtrue
\ifTKDE

\else

\fi

\begin{document}

\title{Efficient Computation of Trip-based Group Nearest Neighbor Queries (Full Version)}

\author{Shahiduz~Zaman,
        Tanzima~Hashem,
        and~Sukarna~Barua
\thanks{S.Zaman ( email: 1018052058@grad.cse.buet.ac.bd), T.Hashem (email: tanzimahashem@cse.buet.ac.bd) and S.Barua (email: sukarnabarua@cse.buet.ac.bd) are with Department of Computer Science and Engineering, Bangladesh University of Engineering and Technology, Dhaka, Bangladesh}
}
\maketitle

\begin{abstract}
In recent years, organizing group meetups for entertainment or other necessities has gained significant importance, especially given the busy nature of daily schedules. People often combine multiple activities, such as dropping kids off at school, commuting to work, and grocery shopping, while seeking opportunities to meet others. To address this need, we propose a novel query type, the Trip-based Group Nearest Neighbor (T-GNN) query, which identifies the optimal meetup Point of Interest (POI) that aligns with users' existing trips.  An individual trip consists of a sequence of locations, allowing users the flexibility to detour to the meetup POI at any location within the sequence, known as a detour location. Given a set of trips for the users, the query identifies the optimal meetup POI (e.g., restaurants or movie theaters) and detour locations from each user's trip that minimize the total trip overhead distance. The trip overhead distance refers to the additional distance a user must travel to visit the meetup POI before returning to the next location in their trip. The sum of these overhead distances for all users constitutes the total trip overhead distance. The computation time for processing T-GNN queries increases with the number of POIs. To address this, we introduce three techniques to prune the POIs that cannot contribute to the optimal solution, and thus refine the search space. We also develop an efficient approach for processing T-GNN queries in real-time. Extensive experiments validate the performance of the proposed algorithm.
\end{abstract}

\begin{IEEEkeywords}
Road networks, group nearest neighbor, group trip planning
\end{IEEEkeywords}

\section{Introduction} \label{ch:intro}
With the growing popularity and widespread use of map-based services such as Google Maps and Bing Maps, developing efficient solutions for new variants of trip planning queries has become increasingly important. Trip planning~\cite{li2005trip,sharifzadeh2008optimal,chen2008multi,ohsawa2012sequenced} and group trip planning queries~\cite{hashem2013group, hashem2015efficient, ahmadi2015mixed, samrose2015efficient} enable individuals or groups to plan trip activities with convenance by allowing them to visit multiple point-of-interests (POIs) like restaurants, shopping centers, and cinema theaters with the minimum travel distance. Traditional group trip planning queries assume that all group members jointly visit a same set of POIs. However, in real-world scenarios, people often have independent predefined travel plans and may seek to meet at a suitable location while performing their scheduled activities. For example, on weekdays, a person may have a predefined trip that begins by dropping their children at school before heading to the office and later stops at a supermarket on the way home. On weekends, a predefined travel plan can include visiting a shopping center, watching a movie in a theater, and attending a family event. During these personal activities, the group may prefer to meet in a location - such as a restaurant or cinema theater for essential or entertainment purposes - that fits their travel plans and minimizes the overall travel distance. To address this need, in this paper, we introduce a novel variant of trip planning query called the Trip-based Group Nearest Neighbor (T-GNN) query identifies identifies an optimal meetup location considering individual travel plans. 

\begin{figure}[htp] \centering{
\includegraphics[width=.9\linewidth]{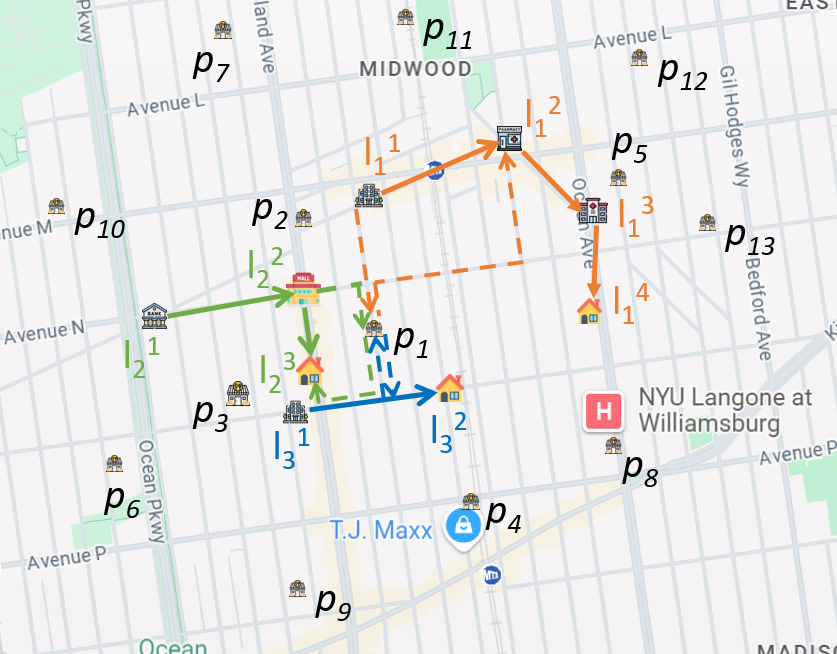}}
\caption{An example scenario for T-GNN query.}
\label{fig:tgnn2}
\end{figure}

Existing research have focused on finding the optimal meetup location for a group without considering personal trip plans. A group nearest neighbor (GNN) query~\cite{abeywickrama2020hierarchical, 
namnandorj2008efficient, papadias2004group, papadias2005aggregate, yiu2005aggregate} finds the optimal location that minimizes the aggregate travel distance of the group members from their source locations. A group trip planning (GTP) query~\cite{hashem2015efficient, hashem2013group, ahmadi2015mixed, samrose2015efficient, li2005trip, mahin2019activity} finds a sequence of locations that minimize the aggregate travel distance of the group members from their source and destination locations via the POIs. In contrast, a T-GNN query finds a meetup location that minimizes the total trip overhead distance of the group considering the individual trip plans of the group members. Each user is assumed to be flexible in making a detour to visit the meetup POI from any location in their predefined trip sequence, which is called the detour location.  After visiting the meetup point, each user resumes their predefined trip from the next location in their sequence. The additional distance a user travels to detour and visit the meetup POI is called the trip overhead distance, and the total trip overhead distance is the summation of trip overhead distances of all the group members. A T-GNN query returns the detour locations along with the optimal meetup POI that minimizes the total trip overhead distance.
Figure~\ref{fig:tgnn2} shows an example scenario of a T-GNN query for a group of three users and a set of 13 POIs $p_1$ to $p_{13}$. User $u_1$ has a predefined trip starting at $l_1^1$, going through $l_1^2$, $l_1^3$, and ending at $l_1^4$. User $u_2$ starts her journey from  $l_2^1$, visits  $l_2^2$, and finally reaches destination $l_2^3$. User $u_3$ travels from  $l_3^1$ to $l_3^2$ to complete her activities. All three users seek a meetup location with respect to their predefined trips that minimizes the total overhead distance. 
The overhead distance (additional) when user $u_1$ detours to POI $p_1$ from $l_1^1$ is $dist(l_1^1, p_1)+dist(p_1,l_1^2)-dist(l_1^1,l_1^2)$. The user returns to her predefined trip at $l_1^2$ after visiting $p_1$. In the figure, the optimal meetup POI for the group is at $p_1$. The detour points are shown in the figure by rectangles, which are $l_1^1$, $l_2^2$, $l_3^1$ for $u_1$, $u_2$, and $u_3$, respectively. These detour points along with $p_1$ minimizes the total trip overhead distance of the users (detoured paths are shown by dotted lines). 


Solving T-GNN queries in real time is computationally challenging. No existing approaches can be applied directly to solve T-GNN queries. Existing GTP algorithm~\cite{hashem2013group} can be adapted to process T-GNN queries as follows. Each pair of consecutive locations in every user's trip is considered as a source-destination pair. A combination of source-destination pairs is formed by selecting randomly one pair per user. For each combination, the GTP algorithm is applied to find an optimal meetup POI, and the final result is the one that minimizes the total trip distance among all combinations. Section~\ref{experiments} discusses the approach in more detail. The number of possible combinations of source-destination pairs grows exponentially with the increase of both the number of users and the number of locations in individual trips. As a result, running GTP algorithms for all possible combinations incurs extremely high processing time, making real-time query processing infeasible.



We develop an efficient algorithm to evaluate T-GNN queries in road networks. Our algorithm incrementally retrieves POIs, and gradually refines the POI search space until the T-GNN quer answer is identified. The key idea behind We develop POI search space refinement prove that 

efficiently in real-time, we propose a solution that explores POIs incrementally from a geometric centroid point and evaluates each POI as an optimal meetup location. As each POI is retrieved, the total overhead distance is calculated with respect to the users' predefined trips, and the algorithm keeps track of the best POI and detour locations from among the POIs already retrieved. The approach terminates when the search process evaluates all the candidate POIs. The query processing overhead is primarily influenced by the number of candidate POIs explored. The key essence of our approach is to significantly reduce the candidate POI set by using three efficient pruning techniques, which substantially lowers the computation time for T-GNN queries.


The main contributions of this paper are as follows:
\begin{enumerate}[label=\roman*.]
    \item We introduce and formally define a novel query, the Trip-based Group Nearest Neighbor (T-GNN) query, which identifies the optimal meetup location while considering users' predefined trips in road networks.
    \item We develop efficient pruning techniques that significantly reduce the search space of candidate POIs explored by our algorithm, thus enhancing the computational efficiency of the T-GNN query processing algorithm.
    \item We propose an efficient approach to find the answer of T-GNN query from the set of candidate POIs in the refined search space.
    \item We perform extensive experiments to validate the performance of the proposed approach using real datasets.
\end{enumerate}


\section{Problem Formulation} \label{problemformulation}
In a T-GNN query, group members have their own predefined trips and seek to organize a meetup event at a location (i.e., a POI) that minimizes the total trip overhead distance of the group.

Formally, given a set of trips (i.e., sequence of locations) for users in the group and a set of POIs, the T-GNN query returns the location of a meetup POI, and for each user, a detour location from the sequence of locations of their trip that minimize the total trip overhead distance. 

A T-GNN query has the following properties.

\begin{enumerate}[label=\roman*.]
    \item A user completes her trip independently maintaining the specified location sequences.  
    \item A user is flexible to visit the meetup POI from any location of their trip. Hence, the user detours from the location that provides the minimum overhead distance. 
    \item A user travels from a selected detour location in their trip to the designated meetup POI and then returns to the next location in their trip.
    \item A user is flexible to adjust the time of their trip so that the user reaches the meetup POI with other users at the similar time.
\end{enumerate}


Let $U = \{u_1, u_2, \dots, u_n\}$ be the set of $n$ users where $n$ is the group size, $\mathcal{L} = \{L_1, L_2, \dots, L_n\}$ be the set of $n$ predefined trips of the users, and $P = \{p_1, p_2, \dots, p_q\}$ be the set of $q$ POIs in the POI database. Each predefined trip $L_i=\{l_i^1, l_i^2,\dots,l_i^{m_i}\}$ is an ordered sequence of locations to be visited by user $u_i$, where $l_i^j$ is the $j$-th location in the trip and $m_i$ is the number of locations. Given two locations $a$ and $b$,  let the function $d_E(a,b)$ return the Euclidean distance and $d(a, b)$ return the shortest path distance in a road network between $a$ and $b$. 

\textbf{A detour location} $l_i^d$ is a location in the user $u_i$'s predefined trip, i.e.,  ($l_i^d \in L_i$).  The user $u_i$ travels to the meetup POI from the detour location $l_i^d$ and then returns to their predefined trip at the immediate next location after $l_i^d$ in $L_i$.

\textbf{Trip distance $T_i$} represents the travel distance of user $u_i$ for traversing the path $l_i^1 \to l_i^2 \to \dots \to l_i^{m_i}$ in their predefined trip $L_i$. In a road network, $T_i$ is calculated as $T_i=d(l_i^1, l_i^2)+d(l_i^2, l_i^3)+\dots+d(l_i^{m_i-1}, l_i^{m_i})$. 

\textbf{Detoured trip distance $T_i(p_k)$} represents the minimum travel distance for user $u_i$, when $u_i$'s predefined trip includes a detour via a meetup POI $p_k$. In the detoured trip, $u_i$ visits the meetup POI $p_k$ in addition to their original locations in the predefined trip. Assuming that $l_i^j \in L_i$ ($j<m_i$) is the detour location that results in the minimum travel distance for the user to visit POI $p_k$, the detoured trip via $p_k$ is $L_i^d=\{l_i^1, l_i^2,\dots,l_i^j, p_k, l_i^{j+1},\dots,l_i^{m_i}\}$. The detoured trip path traversed by the user is $l_i^1 \to l_i^2 \to \dots \to l_i^j \to p_k \to l_i^{j+1}\to \dots \to l_i^{m_i}$.  In a road network, the detoured trip distance $T_i(p_k$) is calculated as:  $T_i(p_k)=d(l_i^1, l_i^2)+d(l_i^2, l_i^3)+\dots+d(l_i^{j-1}, l_i^{j})+
d(l_i^{j}, p_k)+d(p_k, l_i^{j+1})+
d(l_i^{j}, l_i^{j+1})+\dots+d(l_i^{m_i-1}, l_i^{m_i})$. 

\textbf{Trip overhead distance $TO_i(p_k)$} represents the additional distance that a user $u_i$ needs to travel due to the detour via a meetup point $p_k$. Hence, $TO_i(p_k)$ is calculated as: $TO_i(p_k) = T_i(p_k) - T_i=d(l_i^j,p_k)+d(p_k,l_i^{j+1})-d(l_i^j,l_i^{j+1})$, where $l_i^j$ is the detour location that results in the minimum travel distance for the user to visit POI $p_k$.

\textbf{Total trip overhead distance $TO(p_k)$} for a group is defined as: $TO(p_k) = \sum_{i=1}^{n} TO_i(p_k)$, where $p_k$ is the meetup POI for the group.

A T-GNN query for a group of users is formally defined as follows:

\textit{\textbf{Definition 1: A T-GNN query}}. Given a set of  users $U = \{u_1, u_2, \dots, u_n\}$,  a corresponding set of predefined trips $\mathcal{L} = \{L_1, L_2, \dots, L_n\}$, and the set of POIs $P = \{p_1, p_2, \dots, p_q\}$, a T-GNN query returns a meetup POI $p_k$ and $n$ detour locations $l_{1}^{d},l_{2}^{d}, \dots,l_{n}^{d}$, where $l_i^d \in L_i$ for all users, such that the total trip overhead distance $TO(p_k)$ is minimized, i.e., $TO(p_k)\le TO(p_j)$ for every POI $p_j \in (P - \{p_k\})$.  

The notations are summarized  in Table \ref{tab:notion}.

\begin{table}[h!]
  \begin{center}
    \caption{Notations and their meanings}
    \label{tab:notion}
	\begin{tabular}{|p{2.5cm}|p{5cm}|}
	\hline
      \textbf{Notation} & \textbf{Meaning}\\
      \hline
$U=\{u_1, u_2, \dots,u_n\}$ & A set of $n$ users, where $n$ is the group size\\
	\hline
      $P=\{p_1, p_2, \dots,p_q\}$ & A set of $q$ POIs, i.e., $P$ is the POI database\\
	\hline
$m_1, m_2, ..., m_n$& Each $m_i$ denotes the trip length, i.e., number of locations, in the predefined trip of user $u_i$\\
	\hline

$\mathcal{L}=\{L_1, L_2, \dots,L_n\}$& A set of predefined trips of $n$ users\\
	\hline
$L_i=\{l_i^1, l_i^2, \dots,l_i^{m_i}\}$ & An ordered sequence of locations in the trip of user $u_i$\\
\hline
$T_i$ & Trip distance of user $u_i$ without visiting any meetup POI\\
	\hline
$T_i(p_k)$ & Detoured trip distance of user $u_i$ for visiting meetup POI $p_k$\\
	\hline
$TO_i(p_k)$ & Trip overhead distance of user $u_i$, when user detours to visit POI $p_k$\\
	\hline

$TO(p_k)$ &Total trip overhead distance of the group for meeting at POI $p_k$\\
	\hline
$d(a, b)$ & Shortest distance in road network between two points $a, b$\\
	\hline
$d_E(a, b)$ & Euclidean distance between  two points $a, b$\\
	\hline
$l_1^d, l_2^d, \dots,l_n^d$& $n$ detour locations for $n$ users, where detour location of user $u_i$ is $l_i^d$  and $l_i^d \in L_i$\\
	\hline

    \end{tabular}
  \end{center}
\end{table}


\section{Related Works}\label{ch:related}

There exists a substantial works in the literature addressing the group nearest neighbor, trip planning, and group trip planning problems in spatial databases.

\emph{Group Nearest Neighbor Queries.}\label{ch:gnn}
Group nearest neighbor (GNN) queries or more generally aggregate nearest neighbor (ANN) queries~\cite{papadias2004group,papadias2005aggregate, yiu2005aggregate, li2005two, namnandorj2008efficient, abeywickrama2020hierarchical, hashem2019protecting, chung2022efficient} aim to identify a point of interest (POI) that minimizes the aggregate travel distance for a group of users. \emph{Papadias et al.} \cite{papadias2004group} presented three algorithms for GNN queries that minimize total travel distance in Euclidean space. The authors in~\cite{papadias2005aggregate} proposed a GNN query variant that minimizes either the maximum or minimum distance any user must travel. \emph{Yiu et al.}~\cite{yiu2005aggregate} developed R-tree index based algorithms for processing GNN queries on road networks. \emph{Li et al.}~\cite{li2005two} proposed an efficient node pruning technique using elliptical properties for ANN query processing. \emph{Namnandorj et al.}~\cite{namnandorj2008efficient} utilized vector space property to design efficient bounds and proposed  indexed and non-index algorithms for processing ANN queries. \emph{Abeywickrama et al.}~\cite{abeywickrama2020hierarchical} proposed a hierarchical graph traversal approach to find the top-$k$ POIs with the smallest aggregate distances. \emph{Hashem et al.}~\cite{hashem2019protecting} addressed privacy concerns by introducing decentralized techniques for privacy-preserving GNN queries. 

\emph{Trip Planning Queries.}\label{ch:tpq}
Given a set of POIs, a trip planning query (TPQ)~\cite{li2005trip,sharifzadeh2008optimal,chen2008multi,ohsawa2012sequenced, islam2023crowd} returns the shortest trip for a single user, starting from a source, visiting user-specified types of locations (e.g., restaurants, gas stations), and ending at a destination. Extensive research has focused on TPQs for individual users. \emph{Li et al}~\cite{li2005trip} designed a fast approximation algorithm to process TPQs. 
\emph{Sharifzadeh et al.}~\cite{sharifzadeh2008optimal} introduced a TPQ variant known as optimal sequenced route (OSR) query that finds the shortest trip where the user visits data points in a specific sequence of types defined by the user. \emph{Chen et al.}~\cite{chen2008multi} proposed multi-rule partial sequenced route (MRPSR) query and demonstrated that TPQs can be evaluated by the uniform framework provided by MRPSR. Multi-criteria route planning queries~\cite{salgado2018efficient,nadi2011multi} find the optimal route based on multiple factors and accounts for avoiding obstacles such as rivers or private properties. In~\cite{van2008planning, anwar2017optimal,  shen2020euclidean, lambert2003safe}, the authors proposed techniques to compute safe paths~\cite{van2008planning,lambert2003safe} considering localization uncertainties and optimal sequenced route in the presence of obstacles~\cite{anwar2017optimal}. 

\emph{Group Trip Planning Queries.}\label{ch:gtp}
Given a set of users' source-destination pairs and required POI types (e.g., restaurants, shopping centers), a Group Trip Planning (GTP)~\cite{hashem2013group} query returns a meetup point for each POI type that minimize the group's total travel distance. \emph{Hashem et al.}~\cite{hashem2015efficient} developed incremental and dynamic programming algorithms for processing GTP queries in road networks. In~\cite{ahmadi2015mixed, samrose2015efficient}, the authors introduced and proposed solutions for the sequenced GTP query where POI types must be visited in a predefined sequence. 
Tabassum et al.~\cite{TabassumDGTP} proposed dynamic group trip planning (DGTP) queries where the group size can change at any POI during the actual trip. The authors proposed algorithms based on elliptical pruning of POIs to process DGTP queries in real-time. A related variant, known as the Group Trip Scheduling (GTS) query, assigns one trip to each of the $n$ users such that each POI type is visited by exactly one user and the group's total travel distance is minimized. \emph{Jahan et al.}~\cite{jahan2017scheduling, jahan2019efficient} developed algorithms for processing GTS queries in both Euclidean space and road networks, including a dynamic programming technique that eliminates non-contributing trips to improve processing time. \emph{Rayhan et al.}~\cite{rayhan2019efficient} further generalized the GTS problem by proposing Generalized Group Trip Scheduling (GGTS) queries. Unlike GTS, a POI type may be visited by multiple users in a CGTS query. The authors proposed two heuristic methods—Trip Scheduling Heuristic (TSH) and Search Region Refinement Heuristic (SRH)—to efficiently process GGTS queries in real-time.

Our work is closely related to GTP queries with key differences. In GTP queries, users specify only a source-destination pair, and the objective is to identify a sequence of POI types that are visited collectively by all group members. In contrast, a T-GNN query allows each user to provide an independent trip plan, defined as a sequence of locations. The goal is to identify a single meetup POI that is visited by all users. Additionally, T-GNN query lso determines the detour point for each user, indicating where they should deviate from their original trip to reach the meetup location.

\section{Our Approach}{\label{process_tgnn}}
Our approach designates a coordinator from among the group members, while a location-based service provider (LSP) processes T-GNN queries. The POIs are indexed using a $R^*$-tree [REF] in the LSP's database. Each user in the group submits their predefined trip to the coordinator, who then forwards them to the LSP. Upon receiving the predefined trips of $n$ users, the LSP retrieves candidate POIs from the database, computes the optimal solution, and returns the results to the coordinator. The optimal solution consists of a meetup POI and $n$ detour locations, one for each user. The coordinator then communicates the meetup POI and detour locations the respective users. To efficiently process T-GNN queries, our algorithm utilizes the concepts of known areas and search areas, which are determined based on centroid properties of the predefined trip locations, as detailed below.

\textbf{Local and global centroids}. A local centroid is a geometric centroid of a user's predefined trip locations. Let $c_1, c_2, \dots,c_n$ denote the $n$ local centroids for $n$ usersin the group. The local centroid $c_i$ in trip $L_i$ for user $u_i$ is computed as $c_i=\frac{1}{m_i}\sum_{j=1}^{m_i}{l_i^j}$.  The global centroid is the geometric centroid of all local centroids. Let $c_K$ be the global centroid of $c_1, c_2, \dots, c_n$, which is computed as $c_K=\frac{1}{n}\sum_{j=1}^{n}{c_i}$. 

\textbf{Known area}. The known area is a \textit{circular region} with its center at the global centroid $c_K$. To process T-GNN queries, our approach incrementally retrieves nearest POIs from the database based on their distances from $c_K$. The radius of the known area is dynamically updated as the distance from $c_K$ to the most recently retrieved nearest POI. 


\textbf{Search area}. The search area represents the region that contains the candidate POIs for optimal solution computation, though not all of these POIs may have been explored by the algorithm yet. The search area is determined to restrict the POI search space to a significantly smaller region than the entire POI database. 

\begin{figure}\centering{
   \includegraphics[width=.9\linewidth]{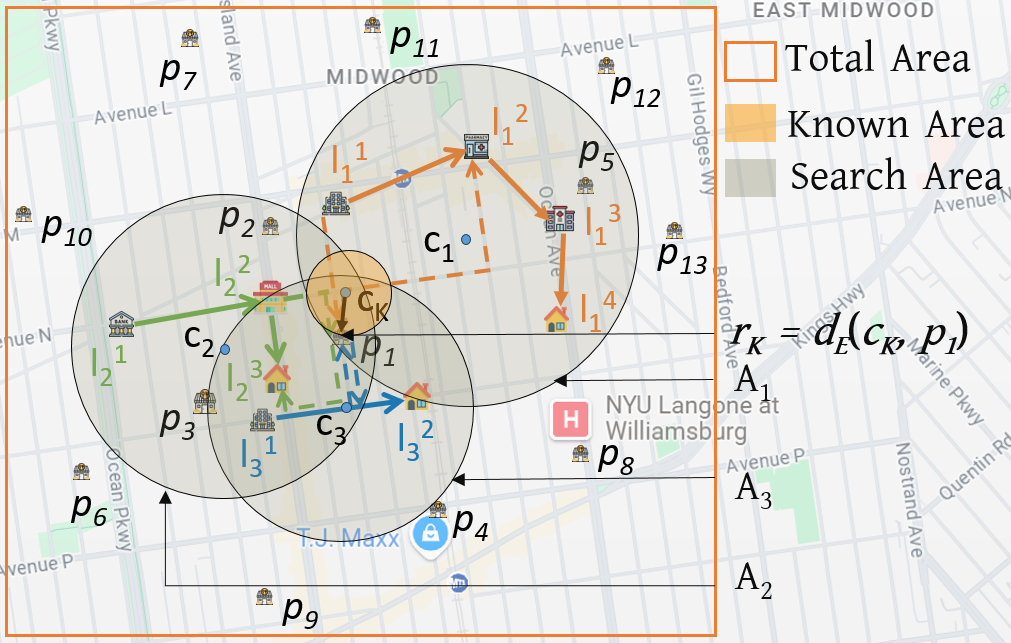}}
  \caption{Known area and search area illustration for the proposed approach.}
  \label{fig:known_search_area}
\end{figure}

Figure~\ref{fig:known_search_area} illustrates the concept of known and search areas for the T-GNN query scenario shown in Figure~\ref{fig:tgnn2}. The local and global centroids are shown in the figure. Suppose our approach retrieves the first POI $p_1$ from the database. The radius of the known area becomes $d_E(c_K,p_1)$, which is the euclidean distance between the global centroid $c_K$ and $p_1$. The union of three circular regions $A_1$, $A_2$, and $A_3$, which are centered on local centroids and have radii determined from the current optimal solution, represents an example search area. The details of the search area computation is discussed in Section~\ref{search_area_refinement}. POIs located outside the search area are guaranteed not be a candidate for optimal solution and can therefore be pruned from further computation.\\

\textbf{Approach overview.}  Our approach incrementally retrieves the nearest POIs from the centroid of the known area $c_K$. After retrieving a next nearest POI $p$, the algorithm updates the known area radius and computes the total trip overhead distance. If the $p$ results in a better solution, the current optimal solution is updated, and search area is refined. The algorithm then checks if the known area covers the search area. If yes, then the algorithm terminates, otherwise it continues to retrieve the next candidate POI from the database. With the retrieval of POIs, known area expands and the computed solution improves, which in turn leads to a reduction of the search area. The key steps of our algorithm are described in the following sections.
\subsection{Initialization} \label{parameter_initialization}
At the beginning, our algorithm initializes the following key variables: known area, search area, total trip overhead distance, meetup POI, and detour locations. The algorithm computes the local centroids $c_1, c_2, \dots, c_n$ and global centroid $c_K$. Initially, the known area is empty and the search area is the entire POI database on the road network. Thus, the radius of the known area is initialized to zero, while the radius of the search area is initialized to infinity. The meetup POI and detour locations are initially empty, and the total trip overhead distance is set to infinity.

\subsection{Retrieval of Candidate POI and Known Area Expansion} \label{known_area_computation}


The algorithm retrieves the next closest candidate POI from the database according to their distance from $c_K$. The algorithm then computes the total trip overhead distance for the candidate POI $p$, and updates the current optimal solution including the meetup POI and detour locations if $p$ results in a better total trip overhead distance. Subsequently, the radius of the known area is updated to the value $d_E(c_k,p)$. This ensures that all candidate POIs that have already been retrieved are contained inside the known area circle. As POIs are retrieved iteratively, the radius of the know area progressively expands. 
In the Figure~\ref{fig:known_search_area}, the POI $p_1$ is the nearest POI to $c_K$ and the algorithm retrieves it first. After retrieving $p_1$, the euclidean distance $d_E(c_k,p_1)$ is set as the radius of the known area. In two subsequent iterations, the algorithm retrieves the next nearest POIs $p_2$ and $p_3$, and updates the radius of the known area to $d_E(c_k,p_2)$ and then to $d_E(c_k,p_3)$. Note that the final known area circle contains all POIs that have already been retrieved from the database (i.e., $p_1$, $p_2$, and $p_3$).

\subsection{Search Area Refinement} \label{search_area_refinement}
The search area is the refined search space encompassing all candidate POIs that the algorithm must explore to compute the optimal solution. The search region is initially set to the entire road network. As POIs are explored by the algorithm, the current optimal solution is updated. The improved solution is utilized to calculate a reduced search space by pruning POIs from the candidate set. 

We present three different pruning techniques for refining the search space of T-GNN query processing algorithm. The techniques build circular regions by utilizing the local centroids and current trip overhead distances. The combination of these circular regions are leveraged as the refined search area.


\subsubsection{Pruning Technique 1 (PT1)} \label{pruning_1}

The search area can be refined as the \textbf{union} of $n$ circular regions $A_1, A_2, \dots, A_n$, where each $A_i$ is a circle having center at $c_i$ and radius as $r_i$ where the value of $r_i$ is defined as $\frac{TO(p)}{2n} + \frac{\max_{1 \leq j  \leq m_i}  d(l_i^j, l_i^{j+1})}{2} +  \max_{1 \leq j  \leq m_i} d(c_i, l_i^j)$. Theorem \ref{theorem_1} and its accompanying proof substantiates this assertion.

\begin{theorem}\label{theorem_1}
Let POI $p$ be the current optimal meetup point having the current minimum total trip overhead distance $TO(p)$. The combined region of $n$ circles $A_1 \cup A_2\cup \dots \cup A_n$ can be used as the refined search space, where each $A_i$ is a circle with center at $c_i$ and radius $r_i$ equal to $\frac{TO(p)}{2n} + \frac{\max_{1 \leq j  \leq m_i}  d(l_i^j, l_i^{j+1})}{2} +  \max_{1 \leq j  \leq m_i} d(c_i, l_i^j)$.
\end{theorem}

\textit{Proof.} Suppose POI $p'$ resides outside the combined area $A_1 \cup A_2\cup \dots \cup A_n$ and using $p'$ as the meetup point leads to a better solution. Let, $l_i^k$ be the detour location for user $u_i$ with respect to meetup point $p'$. Thus, user $u_i$ visits $p'$ from $l_i^k$ and then returns back to $l_i^{k+1}$ after visiting $p'$. Note that $l_i^k$ and $l_i^{k+1}$ are consecutive locations in the user's trip $L_i$. 
since $p'$ resides outside the combined region $A_1 \cup A_2\cup \dots \cup A_n$, $p'$ must also reside outside the circle $A_i$ for user $u_i$.\\

The trip overhead distance for user $u_i$ though $p'$ is
{\small
\begin{equation}
TO_i(p')=d(l_i^k, p') + d(p', l_i^{k+1}) - d(l_i^k, l_i^{k+1})
\label{eq:eq_th1_0} 
\end{equation}
}
Using the triangle inequality property, we have:
{\small
\begin{align}
d_E(c_i, l_i^k) + d_E(l_i^k, p') &\ge d_E(c_i, p') \label{eq:eq_th1_1} \\
d_E(c_i, l_i^{k+1}) + d_E(l_i^{k+1}, p') &\ge d_E(c_i, p') \label{eq:eq_th1_2}
\end{align}
}

\vspace{-5mm}
\begin{equation}
\hspace{-22mm}\text{As } p' \text{ resides outside } A_i, \quad d_E(c_i, p') > r_i
\label{eq:eq_th1_2_1}
\end{equation}
Combining (\ref{eq:eq_th1_1}), (\ref{eq:eq_th1_2}) and (\ref{eq:eq_th1_2_1}) and subtracting the term $d(l_i^k, l_i^{k+1})$, we can write:\\
{\small
\resizebox{\linewidth}{!}{
\begin{minipage}{\linewidth}
\begin{align*}
d_E(c_i, l_i^k) + d_E(l_i^k, p') + d_E(c_i, l_i^{k+1}) +& d_E(l_i^{k+1}, p') - d(l_i^k, l_i^{k+1}) \\
&\geq 2d_E(c_i, p') - d(l_i^k, l_i^{k+1})\\
&> 2r_i - d(l_i^k, l_i^{k+1})
\end{align*}
 \end{minipage}
}
}
Considering that road network distances are greater than or equal to the Euclidean distances:\\
{\small
\resizebox{\linewidth}{!}{
\begin{minipage}{\linewidth}
\begin{align*}
d&(c_i, l_i^k) + d(l_i^k, p') + d(c_i, l_i^{k+1}) + d(l_i^{k+1}, p') - d(l_i^k, l_i^{k+1})\\
&> 2r_i - d(l_i^k, l_i^{k+1})\\
&= \frac{TO(p)}{n} + \max_{ 1 \leq j \leq m_i} d(l_i^j, l_i^{j+1}) +2 \times \max_{ 1 \leq j \leq m_i} d(c_i, l_i^j) - d(l_i^k, l_i^{k+1})\\
&\geq \frac{TO(p)}{n} +2 \times \max_{ 1 \leq j \leq m_i} d(c_i, l_i^j)[\text{As $\max_{ 1 \leq j \leq m_i} d(l_i^j, l_i^{j+1}) - d(l_i^k, l_i^{k+1}) \geq 0 $}]\\
d&(l_i^k, p')+ d(l_i^{k+1}, p') - d(l_i^k, l_i^{k+1})> \frac{TO(p)}{n}\\
&\text{[As $2 \times \max_{ 1 \leq j \leq m_i} d(c_i, l_i^j) - d(c_i, l_i^k) - d(c_i, l_i^{k+1}) \geq 0$]}\\
T&O_i(p')> \frac{TO(p)}{n} [\text{Using (\ref{eq:eq_th1_0})}]
\end{align*}
 \end{minipage}
}
}
Thus total trip overhead distance of all users through POI $p’$ is {\small $TO(p') = \sum_i TO_i(p') > n \times \frac{TO(p)}{n} = TO(p)$}, which is a contradiction. 
As $p'$ reside outside of combined circle and the total trip overhead distance of POI $p'$ is greater than the total trip overhead distance of POI $p$, we can conclude that any POI located outside the search area have greater total trip overhead distance than POI $p$. Thus, we can remove all such POIs from computation and only consider the POIs located inside the search area.
\hfill
\subsubsection{Pruning Technique 2 (PT2)} \label{pruning_2}
The search area can be refined as the \textbf{union} of $n$ circular regions $A_1^2, A_2^2, \dots, A_n$, where each $A_i$ is a circle having center at $c_i$ and radius $r_i$ where the value of $r_i$ is determined as $\frac{TO_i(p)}{2} + \frac{\max_{1 \leq j  \leq m_i} d(l_i^j, l_i^{j+1})}{2} +  \max_{1 \leq j  \leq m_i} d(c_i, l_i^j)$. Theorem \ref{theorem_2} and its accompanying proof substantiates this assertion.

\begin{theorem}\label{theorem_2}
Let POI $p$ be the current optimal meetup point having the minimum total trip overhead distance $TO(p)$ and individual minimum trip overhead distances $TO_i(p)$ for every user $u_i$. The combined region of $n$ circles $A_1 \cup A_2\cup \dots \cup A_n$ can be used as the refined search space, where each $A_i$ is a circle with center at $c_i$  radius $r_i$ equal to $\frac{TO_i(p)}{2} + \frac{\max_{1 \leq j  \leq m_i} d(l_i^j, l_i^{j+1})}{2} +  \max_{1 \leq j  \leq m_i} d(c_i, l_i^j)$.

\end{theorem}

\textit{Proof.} Suppose POI $p'$ resides outside the combined area $A_1 \cup A_2\cup \dots \cup A_n$ and choosing $p'$ as the meetup point results in a better solution. Let, $l_i^k$ be the detour location for user $u_i$ with respect to $p'$. Thus, user $u_i$ visits $p'$ from $l_i^k$ and then returns back to $l_i^{k+1}$ after visiting $p'$. Note that $l_i^k$ and $l_i^{k+1}$ are consecutive locations in the user's trip $L_i$. 
Since $p'$ lies outside the combined region $A_1 \cup A_2\cup \dots \cup A_n$, $p'$ must also lie outside the circle $A_i$ for user $u_i$.\\

The trip overhead distance for user $u_i$ though $p'$ is
{\small
\begin{equation}
TO_i(p')=d(l_i^k, p') + d(p', l_i^{k+1}) - d(l_i^k, l_i^{k+1})
\label{eq:eq_th2_0} 
\end{equation}
}

Using the triangle inequality property, we have:
{\small
\begin{align}
d_E(c_i, l_i^k) + d_E(l_i^k, p') \ge d_E(c_i, p') \label{eq:eq_th2_1}  \\
d_E(c_i, l_i^{k+1}) + d_E(l_i^{k+1}, p') \ge d_E(c_i, p') \label{eq:eq_th2_2} 
\end{align}
}

\vspace{-5mm}
\begin{equation}
\hspace{-22mm}\text{As } p' \text{ resides outside } A_i, \quad d_E(c_i, p')>r_i
\label{eq:eq_th2_2_1} 
\end{equation}

Combining (\ref{eq:eq_th2_1}), (\ref{eq:eq_th2_2}) and (\ref{eq:eq_th2_2_1}) and subtracting the term $d(l_i^k, l_i^{k+1})$, we can write:\\
{\small
\resizebox{\linewidth}{!}{
  \begin{minipage}{\linewidth}
\begin{align*}
d_E(c_i, l_i^k) + d_E(l_i^k, p') + d_E(c_i, l_i^{k+1}) +& d_E(l_i^{k+1}, p') - d(l_i^k, l_i^{k+1}) \\
&\geq 2d_E(c_i, p') - d(l_i^k, l_i^{k+1})\\
&> 2r_i - d(l_i^k, l_i^{k+1}) 
\end{align*}
\end{minipage}
}
}

Since road network distances are always greater than or equal to Euclidean distances, we have:\\
{\small
\resizebox{\linewidth}{!}{
  \begin{minipage}{\linewidth}
\begin{align*}
d&(c_i, l_i^k)+ d(l_i^k, p') + d(c_i, l_i^{k+1}) + d(l_i^{k+1}, p') - d(l_i^k, l_i^{k+1})\\
&> 2r_i - d(l_i^k, l_i^{k+1})\\
&= TO_i(p) + \max_{ 1 \leq j \leq m_i} d(l_i^j, l_i^{j+1}) +2 \times \max_{ 1 \leq j \leq m_i} d(c_i, l_i^j) - d(l_i^k, l_i^{k+1})\\
&\geq TO_i(p) +2 \times \max_{ 1 \leq j \leq m_i} d(c_i, l_i^j)[\text{As $\max_{ 1 \leq j \leq m_i} d(l_i^j, l_i^{j+1}) - d(l_i^k, l_i^{k+1}) \geq 0 $}]\\
d&(l_i^k, p')+ d(l_i^{k+1}, p') - d(l_i^k, l_i^{k+1})> TO_i(p)\\
&\text{[As $2 \times \max_{ 1 \leq j \leq m_i} d(c_i, l_i^j) - d(c_i, l_i^k) - d(c_i, l_i^{k+1}) \geq 0$]}\\
T&O_i(p')> TO_i(p) [\text{Using (\ref{eq:eq_th2_0})}]
\end{align*}
\end{minipage}
}
}

The total trip overhead distance through POI $p’$ is {\small $TO(p') = \sum_i TO_i(p') > \sum_i TO_i(p) = TO(p)$}, which is a contradiction. Since $p'$ resides outside the combined region $A_1 \cup A_2\cup \dots \cup A_n$ and the total trip overhead distance for $p'$ is greater than that of current optimal POI $p$,  we can conclude that any POI located outside the combined region will have a greater total trip overhead distance than $p$.  Hence, we can restrict our search area to $A_1 \cup A_2\cup \dots \cup A_n$, removing all POIs outside this region and considering only those located inside it. 



\subsubsection{Pruning Technique 3 (PT3)} \label{pruning_3}
The search area can be refined as the \textbf{intersection} of $n$ circular regions $A_1, A_2, \dots, A_n$, where each $A_i$ is a circle having center at $c_i$ and radius $r_i$. The value of $r_i$ is defined as $r_i=TO(p)+T_i$, where $TO(p)$ is the  total trip overhead distance with respect to the current optimal meetup point $p$ and $T_i$ is the trip distance of user $u_i$ without visiting any POI. Theorem \ref{theorem_3} and its accompanying proof substantiates this assertion.

\begin{theorem}\label{theorem_3}
Let POI $p$ be the current optimal meetup point having the minimum total trip overhead distance $TO(p)$. The intersection region of $n$ circles $A_1 \cap A_2\cap \dots \cap A_n$ can be used as the refined search area where each $A_i$ is a circle with center at $c_i$ and radius $r_i$ equal to $TO(p)+T_i$.
\end{theorem}
\textit{Proof.} Suppose POI $p'$ resides outside the intersection area $A_1 \cap A_2 \cap \dots \cap A_n$ and using $p'$ as the meetup point results in a better solution. Let, $l_i^k$ be the detour location for user $u_i$ with respect to $p'$. Thus, user $u_i$ visits $p'$ from $l_i^k$ and then returns back to $l_i^{k+1}$ after visiting $p'$. Note that $l_i^k$ and $l_i^{k+1}$ are consecutive locations in the user's trip $L_i$. 
since $p'$ resides outside the combined region $A_1 \cap A_2 \cap \dots \cap A_n$, $p'$ must also reside outside the circle $A_i$ for user $u_i$.  The trip overhead distance of user $u_i$ though $p'$ is $TO_i(p')$.

As $p'$ resides outside the circle $A_i$, we have\\
\centerline{$d_E(c_i, p') >r_i$}\\
In the 5.3 from Hashem et al.~\cite{hashem2015efficient} shows that distance between a POI and a centre of n locations is less than or equal to the average distance of the POI from the n locations set. If $c_i$ is the geometric centroid of user $u_i$’s trip locations $L_i=\{l_i^1, l_i^2, \dots, l_i^{m_i}\}$, the following inequality holds for any POI $p$, where $d_E$ denotes the euclidean distance.

{\small
\begin{equation}
d_E(c_i, p) \leq \frac{1}{m_i}\sum_{1 \leq j \leq m_i} d_E(l_i^j, p)
  \label{eq:lemma_1} 
\end{equation}
}
From equation~\ref{eq:lemma_1} we have, $d_E(c_i, p') \leq \frac{1}{m_i}\sum_{1 \leq j \leq m_i} d_E(l_i^j, p')$. There is at least one location $l_i^r$ in $L_i$ such that\\
{\small
\begin{equation}
d_E(l_i^r, p') \geq d_E(c_i, p') > r_i
  \label{eq:eq1} 
\end{equation} 
}

If $l_i^r$ is the detour location for $u_i$, then $d(l_i^r, p')$ is a part of $T_i(p')$, In this case, we have\\
{\small
\begin{equation}
T_i(p') \geq d(l_i^r, p') \geq d_E(l_i^r, p')>r_i\\
\label{eq:eq2} 
\end{equation} 
}

If $l_i^r$ is not the detour location for $u_i$, then let $l_i^t$ be the detour location. The following two cases may happen.

Case 1: $r<t$, i.e., user travels through $l_i^r$, then $l_i^t$, and then $p'$. In this case, we have, $T_i(p') \geq d(l_i^r,l_i^t) + d(l_i^t,p')$ [as both are parts of $T_i (p')$ where $d(l_i^r,l_i^t )$ is the partial trip distance from $l_i^r$ to $l_i^ t$]. Since road network distances are greater than or equal to Euclidean distances, we have 

{\small
\begin{equation}
T_i(p') \geq d(l_i^r,l_i^t) + d(l_i^t,p') \geq d_E(l_i^r, l_i^t) + d_E(l_i^t, p')
\label{eq:eq3} 
\end{equation} 
}
Using triangle inequality, we have: 
{\small
 \begin{equation}
T_i(p') \geq d_E(l_i^r, l_i^t) + d_E(l_i^t, p') \geq d_E (l_i^r, p')>r_i
\label{eq:eq4} 
\end{equation} 
}

Case 2: $t<r$, i.e., user travels to $l_i^t$, then $p'$, and then $l_i^r$. If $r=t+1$, i.e., user immediately visits $l_i^r$ after $p'$, then $d(l_i^r,p')$ is a part of $T_i(p')$ and we get $T_i(p')\geq d(l_i^r,p')\geq d_E(l_i^r,p')>r_i$. Otherwise, user visits $l_i^{t+1}$ after $p'$ and then visits $l_i^r$. In this case:
$T_i(p') \geq d(l_i^{t+1},p')+d(l_i^{t+1},l_i^r)$ where $d(l_i^{t+1},l_i^r)$ denotes the partial trip distances from $l_i^{t+1}$ to $l_i^r$.
Since road network distances are always greater than or equal to Euclidean distances, we have
{\small
\begin{equation}
T_i(p')  \geq d(l_i^{t+1},p')+d(l_i^{t+1},l_i^r) \geq d_E(l_i^{t+1},p')+d_E(l_i^{t+1},l_i^r)
\label{eq:eq5} 
\end{equation} 
}
Using triangle inequality:
{\small
\begin{equation}
T_i(p') \geq d_E(l_i^{t+1},p')+d_E(l_i^{t+1},l_i^r) \geq d_E(l_i^r,p')>r_i
\label{eq:eq6} 
\end{equation}
}
Thus, in all cases, we get
{\small
\begin{align*}
T_i(p') \geq r_i &= TO(p) + T_i  \text{ [as $r_i=TO(p)+T_i$] }\\
T_i(p') - T_i &\geq TO(p)
\end{align*}
}
As $T_i(p')-T_i$ is the trip overhead distance of $u_i$ through $p'$
{\small
\begin{align*}
TO_i(p') &\geq TO(p)  
\end{align*}
}

Combining for all users, we have:  {\small $TO(p') = \sum_i TO_i(p') \geq TO(p)$}, which is a contradiction. Since, the total trip overhead distance through $p'$ cannot be greater than the total trip overhead distance through $p$ as $p'$ gives us a better solution than $p$.

\textbf{Combination of pruning techniques.}
In PT1 and PT2, the search areas are refined by the combined area of $n$ circular regions, and in PT3, the search area is refined by the intersected area of $n$ circular regions. Let the refined search areas according to PT1, PT2, and PT3 be $S_1$, $S_2$, and $S_3$, respectively. A POI must satisfy all three pruning techniques to be considered as a candidate. Hence, the final search area for T-GNN queries will be $S_1 \cap S_2 \cap S_3$, the intersected area of $S_1$, $S_2$ and $S_3$. Any POI located outside this intersected search area results in a greater total trip overhead distance than the current optimal solution, and thus it is removed from the candidate POI set. \\



\begin{figure*}[t!]
   \begin{center}
        \begin{tabular}{cccc}
            
             \includegraphics[height=3.3cm,width=0.23\textwidth]{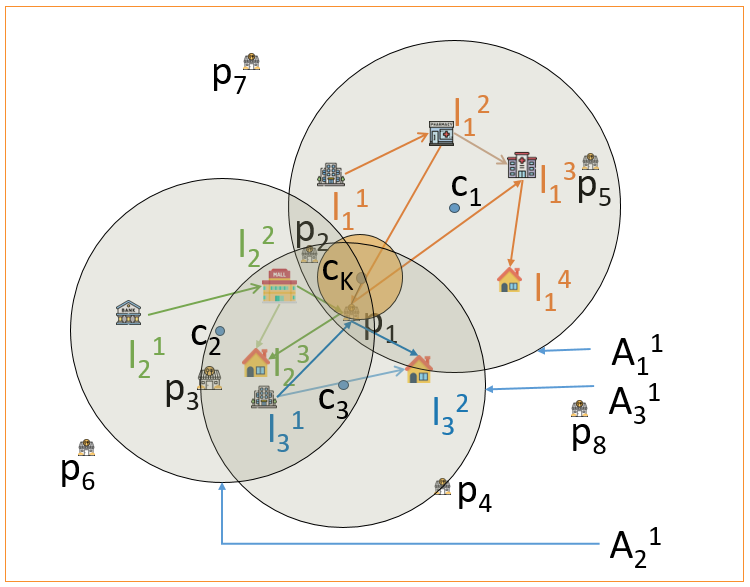}&
             \includegraphics[height=3.3cm,width=0.23\textwidth]{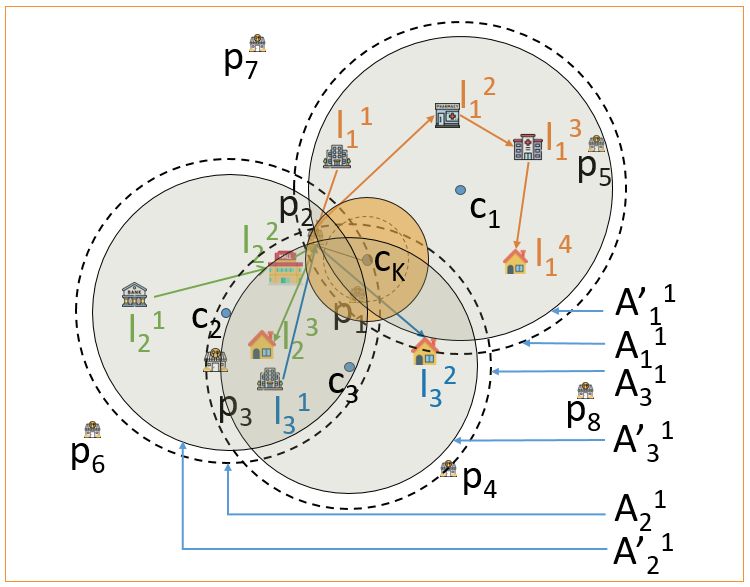}&
             \includegraphics[height=3.3cm,width=0.23\textwidth]{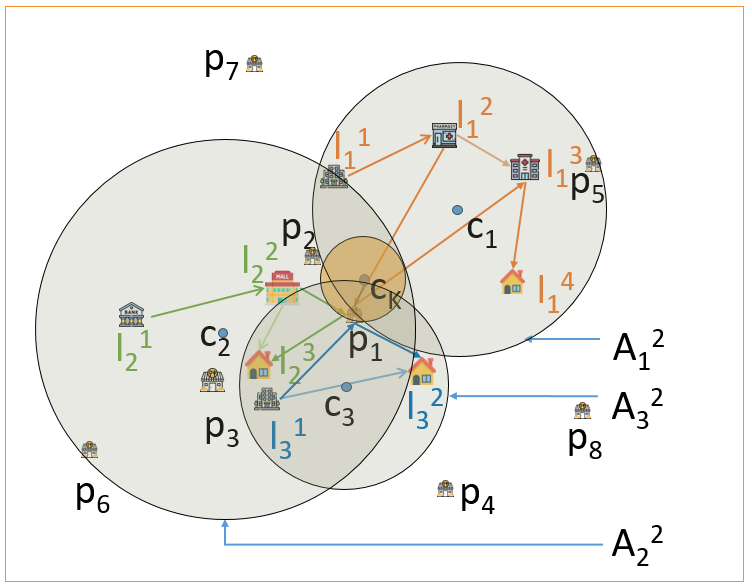}&
             \includegraphics[height=3.3cm,width=0.23\textwidth]{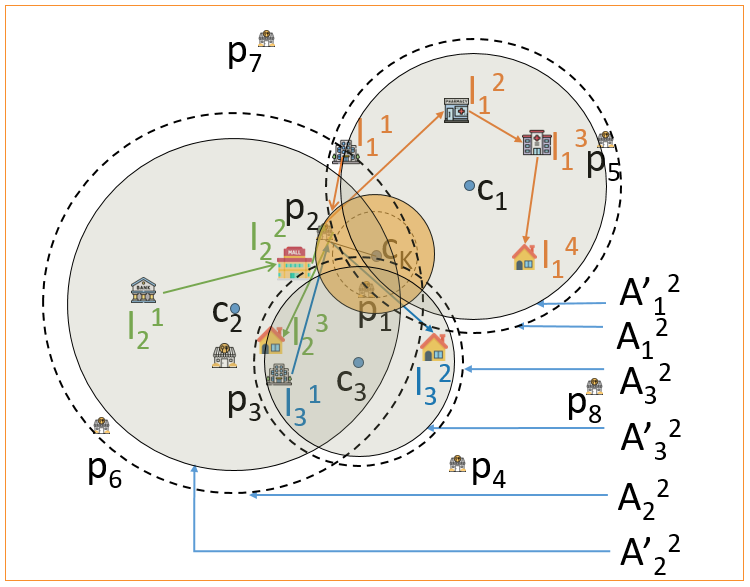}\\
             \scriptsize{\parbox{0.23\textwidth}{(a) Search area for PT1 after $p_1$ is retrieved \textsc{}\hspace{0mm}}}  &  \scriptsize{\parbox{0.23\textwidth}{(b) Refined search area for PT1 after $p_2$ is retrieved \textsc{}\hspace{0mm}}} &  \scriptsize{\parbox{0.23\textwidth}{(c) Search area for PT2 after $p_1$ is retrieved. \textsc{}\hspace{0mm}}} &  \scriptsize{\parbox{0.23\textwidth}{(d) Refined search area for PT2 after $p_2$ is retrieved  \textsc{}\hspace{10mm}}}\\
            \vspace{1mm}\\

             \includegraphics[height=3.3cm,width=0.23\textwidth]{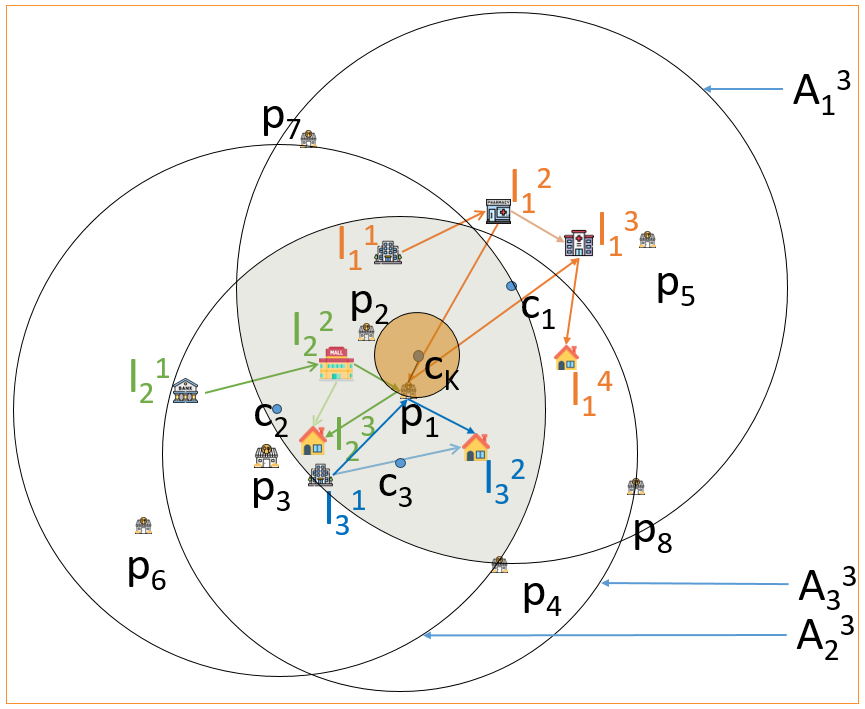}&
             \includegraphics[height=3.3cm,width=0.23\textwidth]{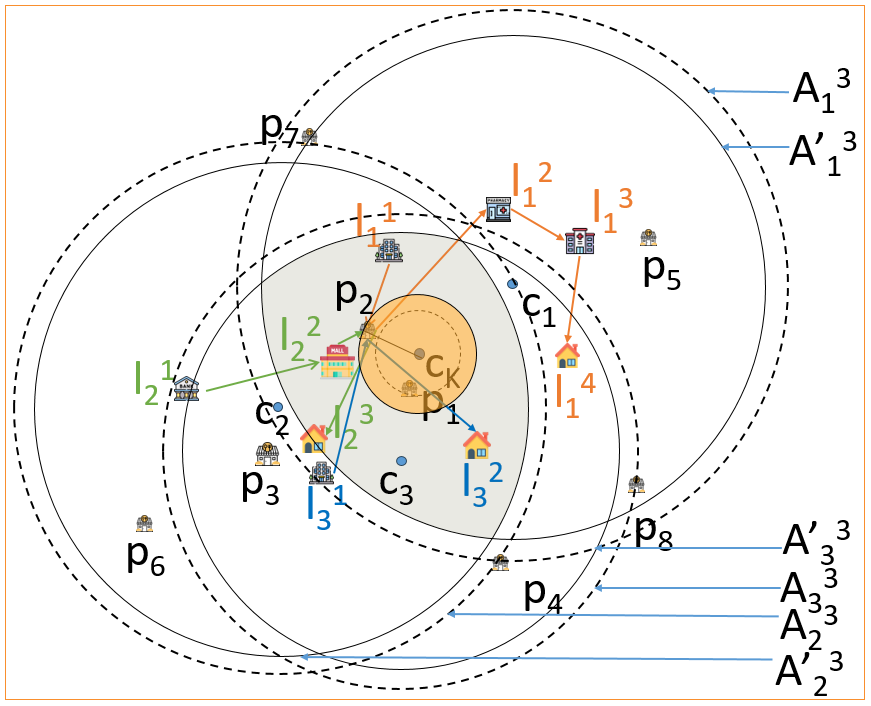}&
             \includegraphics[height=3.3cm, width=0.23\textwidth]{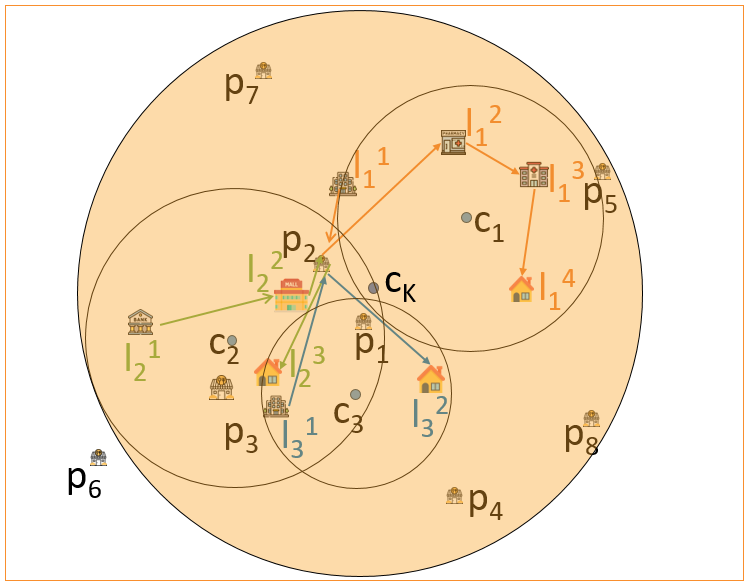}&
             \includegraphics[height=3.3cm,width=0.23\textwidth]{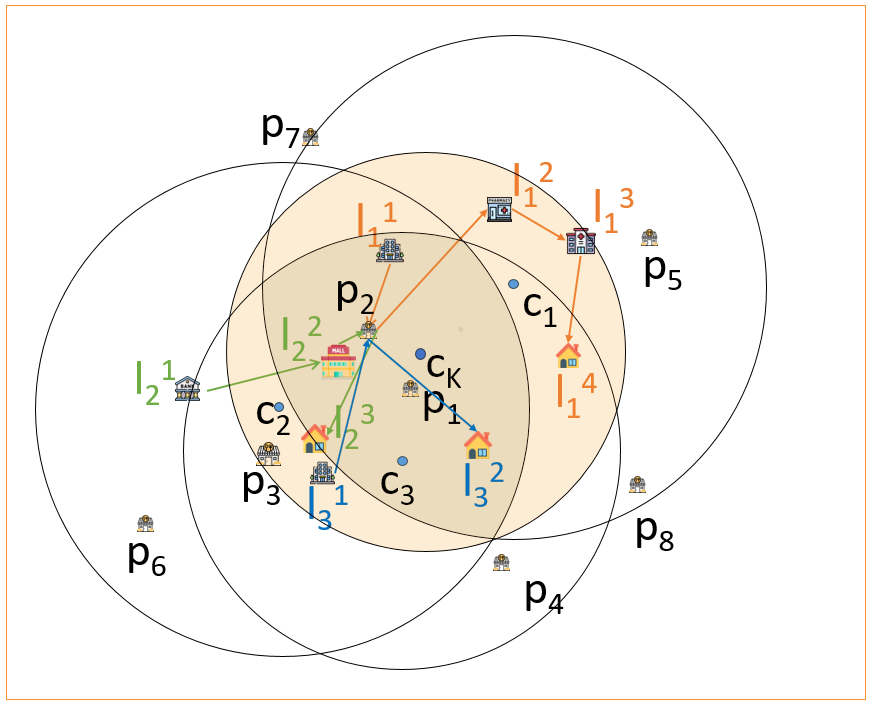}\\ 
             \scriptsize{\parbox{0.23\textwidth}{(e) Search area for PT3 after $p_1$ is retrieved \textsc{}\hspace{0mm}}}  &  \scriptsize{\parbox{0.23\textwidth}{(f) Refined search area for PT3 after $p_2$ is retrieved \textsc{}\hspace{0mm}}} &  \scriptsize{\parbox{0.23\textwidth}{(g) Termination criteria for PT1 and PT2 when known area circle covers the search area \textsc{}\hspace{0mm}}} &   \scriptsize{\parbox{0.23\textwidth}{(h) Termination critera for PT3 when known area covers search area \textsc{}\hspace{10mm}}}\\
            \hspace{-6mm}
        \end{tabular}
       \vspace{-3mm}
        \caption{Illustration of T-GNN Pruning Techniques (a-f) and Termination Criteria (g-h)}
       \vspace{-7mm}
      \label{fig:pruning_steps} 
    \end{center}
\end{figure*}

\subsection{Termination} 
\label{terminating_the_algorithm}
The process of known area expansion and search area refinement continues with the incremental retrieval of POIs. At some point, the known area covers the refined earch area completely, meaning that all the candidate POIs within the search area have been retrieved and explored by the algorithm. The algorithm then terminates and returns the current optimal solution as the final result. 
\subsection{Illustration of search area refinement}


Consider the T-GNN query scenario discussed previously (Figure~\ref{fig:known_search_area}). The first POI $p_1$ is retrieved from the database as it is the closest to $c_K$. The euclidean distance $d_E(c_K,p)$ becomes the new radius of the known area. Figures~\ref{fig:pruning_steps}(a), ~\ref{fig:pruning_steps}(c), and ~\ref{fig:pruning_steps}(e) show the computed search areas for PT1, PT2, and PT3, respectively. In Figure~\ref{fig:pruning_steps}(a), the search area $S_1$ is the union of three circular regions $A_1^1 \cup A_2^1 \cup A_3^1$, shown by gray area in the figure.  
Similarly, Figure~\ref{fig:pruning_steps}(c) shows the computed search area for PT2 where the refined search area $S_2$ is the union of three circular regions $A_1^2 \cup A_2^2 \cup A_3^2$. Finally, Figure~\ref{fig:pruning_steps}(e) displays the computed search area for PT3 where the refined search area $S_3$ is the intersection of the three circular regions $A_1^3 \cap A_2^3 \cap A_3^3$. The final refined search area for T-GNN query is the intersection of $S_1$, $S_2$ and $S_3$.

The next POI retrieved from the database is $p_2$. The radius of known area is updated as $d_E(c_K,p_2)$. Since $p_2$ is inside the combined search area $S_1 \cap S_2 \cap S_3$, it is used for further processing. Assume that the total trip overhead distance through the meetup POI $p_2$ is better than that of $p_1$.  This means the search area will now be updated by our algorithm based on the pruning techniques.

Figure~\ref{fig:pruning_steps}(b) shows the updated search area $S_1$. The circular regions corresponding to the three users are updated and labeled as $A_1^{'1}, A_2^{'1}, A_3^{'1}$ in the figure. The search area $S_1$ becomes $A_1^{'1} \cup A_2^{'1} \cup A_3^{'1}$. Similarly, Figure~\ref{fig:pruning_steps}(d) shows the updated circular regions $A_1^{'2}, A_2^{'2}, A_3^{'2}$ and the latest search area $S_2$ becomes $A_1^{'2} \cup A_2^{'2} \cup A_3^{'2}$. Finally, Figure~\ref{fig:pruning_steps}(f) displays the updated search area $S_3$ where $S_3$ is the intersected area of circular regions $A_1^{'3} \cap A_2^{'3} \cap A_3^{'3}$. 

Figures~\ref{fig:pruning_steps}(g) demonstrates the termination criteria for PT1 and PT2 (refined search area is a union for both pruning techniques) while Figure~\ref{fig:pruning_steps}(h) illustrates the terminating scenario for PT3. The refined search area for T-GNN query becomes $S_1 \cap S_2 \cap S_3$. Thus, our algorithm terminates when the known area covers $S_1 \cap S_2 \cap S_3$.

\section{Algorithm for T-GNN} \label{t_gnn}


\begin{small}
\begin{algorithm}
        \caption{$EA\_T-GNN(\mathcal{L}, root)$}
        \label{alg:algo_1}
        \hspace*{\algorithmicindent} \textbf{Input:} $\{\mathcal{L}, root\}$\\
        \hspace*{\algorithmicindent} \textbf{Output:} $\{R=\{p, L^d\}\}$
        \begin{algorithmic}[1]
            \STATE  $TO_{opt} \gets \infty$, $L^d \gets \phi$
            \STATE $Initialize(\mathcal{L}, c_K, r_K, S_1, S_2, S_3, T, mdist, cdist)$
            
            \STATE $Enqueue(Q, \{root, d_E(c_K, root)\})$
                \WHILE{$Q\ne \phi$  and $Contains(c_K,r_K, S_1, S_2, S_3) = false$}
                	\STATE $\{p , d_E(c_K, p)\}   \gets Dequeue(Q)$
                	\STATE $r_K \gets d_E(c_K, p)$
                	\IF{$p$ is not a POI}
                    		\FOR{$node \in p.children$}               
                    		      \STATE $Enqueue(Q,\{node, d_E(c_K, node)\})$
                    		\ENDFOR
                	\ELSIF{$Prune(p, S_1, S_2, S_3) = false$}	
                            \STATE $TO_{new} \gets 0$, $L^d_{new} \gets \phi$
                            \FOR{$i \gets 1 \ldots n$}
                                \STATE $\{TO[i], l^d_i\} \gets ComputeTO(p, L_i)$
                                \STATE $TO_{new} \gets TO_{new} +TO[i]$, $L^d_{new} \gets L^d_{new} \cup \{l^d_i$\}
                            \ENDFOR
                            \IF{$TO_{new} < TO_{opt}$}
                    			 \STATE $TO_{opt} \gets TO_{new}$, $L^d \gets L^d_{new}$
                    			 \STATE $R \gets \{p, L^d\}$	
                                   \FOR{$i \gets 1 \dots n$}
                                        \STATE $S_1[i].radius \gets \frac{TO_{opt}}{2n} + \frac{mdist[i]}{2} + cdist[i]$
                                        \STATE $S_2[i].radius \gets \frac{TO[i]}{2} + \frac{mdist[i]}{2} + cdist[i]$
                                        \STATE $S_3[i].radius \gets TO_{opt} + T[i]$
                                    \ENDFOR
                            \ENDIF
                        \ENDIF
                \ENDWHILE
          \STATE \textbf{return} $R$
       \end{algorithmic}
    \end{algorithm}
\end{small}
Algorithm~\ref{alg:algo_1}, EA\_T-GNN shows the steps of our approach to process T-GNN queries. All the POIs are indexed using an R-tree in the POI database. The algorithm takes a set of trips $\mathcal{L}=\{L_1, L_2,\dots,L_n\}$ of $n$ users and the root of the R-tree as input. The output of the algorithm is the result set $R$ which contains the meetup POI $p$ and the detour locations $\{l^d_1, l^d_2, \dots, l^d_n\}$. Each user $u_i$ travels to the meetup POI $p$ from the detour location $l^d_i$ where $l^d_i \in L_i$.\\

At the beginning, the total trip overhead distance $TO_{opt}$ is initialized to infinity, and the detour locations $L^d$ is initialized to empty (Line 1). The  $Initialize$ function (Algorithm 2) is used to initialize parameters search areas $S_1, S_2, S_3$, $T$, maximum distance between consecutive locations in each user's trip ($mdist$), and maximum distance between a location in a user's trip and centroid $c_K$ ($cdist$) (Line 2). The root of the R-tree along its euclidean distance to the global centroid $c_K$ is enqueued in the priority queue (Line 3).

\begin{small}
\begin{algorithm}
    \caption{$Initialize(\mathcal{L}, c_K, r_K, S_1, S_2, S_3, T, mdist, cdist)$}
    \label{alg:compute_static_distance}
        \begin{algorithmic}[1]
          \FOR{$i\leftarrow 1 \dots n$}
                \STATE    $c_i\leftarrow \frac{1}{m_i}\sum_{j=1}^{m_i}{l_i^j}$, $mdist[i] \leftarrow 0$, $cdist[i] \leftarrow 0$, $T[i] \leftarrow 0$
                \STATE    $S_1[i].center = c_i, S_2[i].center = c_i, S_3[i].center = c_i$
                \STATE    $S_1[i].radius = \infty, S_2[i].radius = \infty, S_3[i].radius = \infty$
                \FOR{$j\leftarrow 1 \dots m_i$}
                        \IF{$j < m_i$}
                            \STATE $tempDist\leftarrow d(l_i^j,l_i^{j+1})$, $T[i]\leftarrow T[i]+tempDist$
                            \IF{$mdist[i] < tempDist$}
                                \STATE $mdist[i] \leftarrow tempDist$
                            \ENDIF
                        \ENDIF
                        \IF{$cdist[i] < d_E(l_i^j, c_i)$ }
                            \STATE $cdist[i] \leftarrow d_E(l_i^j, c_i)$
                        \ENDIF
                \ENDFOR
            \ENDFOR
            \STATE $r_K\leftarrow 0$, $c_K\leftarrow \frac{1}{n}\sum_{j=1}^{n}{c_j}$
       \end{algorithmic}
    \end{algorithm}
\end{small}

Inside the while loop (Lines 4 - 26), nodes are dequeued on the basis of the smallest distance from $c_K$ (Line 5). The known area radius $r_K$ is expanded to the euclidean distance from $c_K$ to the retrieved node $p$ (Line 6). 

If the dequeued node is not a POI, then we enqueue every child of the dequeued node (Line 9). Otherwise, we check whether the POI can be pruned.


Function $Prune$ checks whether POI $p$ resides inside any of the search areas $S_1, S_2$, and $S_3$ (Line 11). If it is true, then our algorithm processes the POI. The function $ComputeTO$ takes the POI $p$ and trip $L_i$, and computes the minimum trip overhead distance $TO[i]$ and optimal detour location $l^d_i$ for the user $u_i$ with respect to the meetup POI $p$ (Line 14). The function $ComputeTO$ is defined in Algorithm~\ref{alg:algo_2}.

\begin{small}
\begin{algorithm}
    \caption{$ComputeTO(p, L_i)$}
    \label{alg:algo_2}
        \begin{algorithmic}[1]
            \STATE $TO_i \gets \infty$              
	      \FOR{$j \gets 1$ to  $sizeof(L_i)-1$}
			\STATE $OD \gets d(l_i^j, p) + d(l_i^{j+1}, p) - d(l_i^j, l_i^{j+1})$\;
			\IF{ $OD < TO_i$}
    				\STATE $TO_i \gets OD$, $l^d_i \gets l_i^j$
                \ENDIF
            \ENDFOR
            \STATE \textbf{return} \{$TO_i, l^d_i$\}
       \end{algorithmic}
    \end{algorithm}
\end{small}

The trip overhead distances of all users are summed to compute the total trip overhead distance $TO_{new}$ (Line 15). If $TO_{new}$ is better than $TO_{opt}$, the algorithm updates the result set $R$ and total trip overhead distance $TO$ using the new values $TO_{new}$, $L^d_{new}$, and $p$. The algorithm then updates the radius of three search areas $S_1$, $S_2$, and $S_3$ for the pruning technique 1, 2, and 3 using the updated values of $TO_{opt}$ and $TO[i]$s (Lines 21-23).

The function $Contains$ checks whether the known area circle ($c_K$ and $r_K$) covers the intersection of the three search areas $S_1, S_2$, and $S_3$. 
If yes, then all POIs required for optimal solution have been retrieved from the tree. Thus, the algorithm terminates when $Q = \phi$ or $Contains$ function returns true (Line 4).


Algorithm 2 outlines the steps for the initialize function. For each user $u_i$, it computes the local centroid $c_i$, and sets the center and radius of each search area circle to $c_i$ and infinity, respectively (Lines 2-4). The trip distance without visiting any meetup POI ($T[i]$), maximum distance between consecutive locations in the user's trip ($mdist[i]$), and maximum distance between a location in the user's trip and the local centroid $c_i$ ($cdist[i])$ are computed next (Lines 5-16). Finally, the known area radius $c_K$ is initialized to zero, and global centroid $c_K$ is computed (Line 17). 

\subsection{Complexity Analysis}\label{ch:complexity}
Let, $v$ denote the number of vertices, and $e$ the number of edges in the road network. The function $Initialize$ in Algorithm~1 calculates shortest path distances between each pair of adjacent locations for every user's trip (Line 7 of Algorithm~2). This computation is performed using Dijkstra’s algorithm, resulting in an overall time complexity of $O(nm(e+v)\log(v))$ for the $Initialize$ function, where $m$ is the average trip length.

The while loop in Algorithm~1 executes at most $O(q)$ times, where $q$ is the total number of POIs in the database. The most computationally expensive operation within the loop is the $ComputeTO$ function which uses Dijkstra's algorithm to compute trip overhead distances for every user (Lines 13-15 of Algorithm~1). Each call to the $ComputeTO$ function has a time complexity of $O(m(e+v)\log(v))$. This results in a total time complexity of $O(qnm(e+v)\log(v))$ for $n$ users and $q$ iterations of the while loop.

Thus, the overall time complexity of the EA\_T-GNN algorithm (Algorithm~1) is $O(nm(e+v)\log(v) + qnm(e+v)\log(v))$, which simplifies to $O(qnm(e+v)\log(v))$. This indicates that the time complexity of EA\_T-GNN is primarily dominated by the distance computations in the road network, performed via Dijkstra's algorithm.  We do not use all pairs shortest path algorithm for two reasons. First, the space complexity is prohibitively high. Second, the number of trip distances required is significantly smaller than the number of all pairs distances in the road network.

The actual run-time of EA\_T-GNN is significantly reduced by the application of pruning techniques PT1, PT2 and PT3. Let, $r>1$ denote the combined pruning factor, which reduces the number of POIs processed from the queue by a factor of $r$. This reduces the time complexity of EA\_T-GNN algorithm to $O(\frac{q}{r}nm(e+v)\log(v))$.


\section{Experiments}{\label{experiments}}
In this section, we evaluate the proposed approach EA\_T-GNN to process T-GNN queries on real-world datasets. Since there is no existing work to process T-GNN queries, we compare our proposed approach with a baseline algorithm developed by modifying the GTP query processing algorithm~\cite{hashem2013group}. We also demonstrate the effectiveness of different pruning techniques and scalability of the proposed approach. 



\subsection{Baseline Algorithm}\label{ch:baseline}
We design a baseline algorithm BA\_T-GNN for processing T-GNN queries using the group trip planing (GTP) algorithm~\cite{hashem2015efficient}. 
In a GTP query, user's only provide source and destination locations, and GTP algorithm returns a set of POIs that minimize the aggregate travel distance, where all users visit all POIs. The following steps modify GTP algorithm to process T-GNN queries.\\
\begin{adjustwidth}{1cm}{0cm}
\begin{mylist}{Step }
    \item For each user trip, which may include more than two locations, identify all possible source-destination pairs. Each pair represents two consecutive locations in the sequence of the user’s trip.
    \item Compute the set of unique combinations of source and destination pairs, where each combination contains exactly one source-destination pair from each user's set of source-destination pairs.
    \item Apply the GTP algorithm for each combination to find the POI that minimize the total travel distance of the users with respect to a single POI type (i.e., meeting locations). 
    \item Select the POI that provides the minimum total travel distance among all combinations as the answer of T-GNN query. The source locations in the corresponding combination are selected as the detour locations of the users. Note that minimizing the total travel distance for the GTP query is equivalent to minimizing the total trip overhead distance in the T-GNN query.
\end{mylist} 
\end{adjustwidth}
\vspace{0.2cm}
The high level workflow of the GTP algorithm~\cite{hashem2015efficient} is as follows. The algorithm models the known and search areas using ellipses, where the foci of each ellipse are the geometric centroids of the source and destination locations, respectively. It incrementally retrieves POIs from the database based on the summation of the euclidean distance of the POI from both foci of the known area until the GTP query answer is identified. With the retrieval of POIs, the algorithm gradually refines the search space using elliptical properties. The key limitations of this baseline approach are (1) retrieval of the same POIs from the database across multiple GTP combinations and (2) computations of a huge number of road network distances for all combinations.

\subsection{Experiment Setup}\label{ch:experiment_arrangement}
\textbf{Datasets}.
We use real world road network of two cities, Chicago (CH) and New York (NY). The road network datasets are available in OpenStreetMap (OSM) \footnote{ \href{https://www.openstreetmap.org}{https://www.openstreetmap.org}} format. Table~\ref{tab:dataset} shows the the number of nodes and edges available in each dataset. 
We normalize the road network in each dataset to a 
$1000 \times 1000$ square unit space. For the experiments, we randomly generate T-GNN queries as follows. First, we randomly select a rectangular area in the road network as the query area. The trip locations of all users are then generated randomly by choosing network nodes within the query area. The POI dataset is also generated randomly, based on a specified density relative to the total number of road network nodes. For instance, a 10\% density corresponds to 22,722 POIs in the New York dataset. Note that although the users' trip locations are confined to the query area, the final meetup location can be any POI from the entire POI dataset.

\begin{table}[h!]
  \begin{center}
    \caption{Size of real-world datasets}
    \label{tab:dataset}
	\begin{tabular}{ccc}
	\toprule
      Dataset & Nodes & Edges\\
      \hline
       New York (NY) & 227222 & 381021\\
       Chicago (CH) & 134405 & 242357\\
      \bottomrule
    \end{tabular}
  \end{center}
\end{table}

\textbf{Parameters and Measures.}
To evaluate the performance of T-GNN queries, we conduct several experiments by varying the group size $n$ (number of users in the T-GNN query), trip length $m$ (number of locations in the predefined trips), POI density $\rho$ (number of POIs in the database), and query area $QA$. Although $m$ may vary across users, we keep it fixed for brevity in our experiments. Table~\ref{tab:parameter} shows the range of values for the four parameters along with the default values. For each experiment, one parameter is varied while other parameters are set to their default values. 
It was demonstrated that POI density in a real-world network typically remains at around 1\% of the network nodes~\cite{zhong2015g,abeywickrama2016k}. Following this, we set the default POI density$(\rho)$ to 1\%, and it is varied from 0.1\% to 5\%. For other parameters, we reasonably set their values following the combinations used in existing works. 

\begin{table}[h!]
  \begin{center}
    \caption{Parameter Settings}
    \label{tab:parameter}
        \begin{tabular}{p{0.30\linewidth}  p{0.4\linewidth} p{0.13\linewidth}}
	\toprule
      Parameter & Values & Default\\
      \hline
      Group size $(n)$ & 2, 4, 6, 8, 10 & 6\\
      Trip length $(m)$ & 2, 4, 6, 8, 10 & 6\\
      POI density $(\rho)$ & 0.1\%, 0.5\%, 1\%, 5\% & 1\% \\
      Query area $(QA)$ (in sq. units) & $25 \times 25$, $50 \times 50$, $75 \times 75$, $100 \times 100$ & $50 \times 50$ \\
      \bottomrule
    \end{tabular}
  \end{center}
\end{table}

We implement the algorithms in C++. All experiments are conducted on a 64-bit Windows 10 machine having Intel Core i7 CPU and 16 GB as RAM.  The computational time is used as the main performance measure for evaluating the approaches. Besides, we demonstrate the efficiency of the pruning techniques in terms of the number of POIs retrieved. Each experiment is repeated thirty times with random samples, and the average results are reported.

\subsection{Results}\label{ch:performance_evaluation}
In the following subsections, we first present the comparative analysis of our EA\_T-GNN approach with the baseline algorithm. We then present the results showing the effectiveness of our pruning techniques and scalability of EA\_T-GNN.

\subsubsection{Effect of group size}\label{ch:effect_user}
Figures~\ref{fig:varyParams}(a) and~\ref{fig:varyParams}(b) show the effect of group size on the performance of T-GNN query processing algorithms for NY and CH datasets, respectively. We observe that the EA\_T-GNN approach significantly outperforms the baseline method on both the NY and CH datasets. As the number of users increases, the computational time increases for both approaches. However, the computation time for the baseline (BA\_T-GNN) increases exponentially while that of EA\_GNN increases only linearly. For $n=10$, the computation time of BA\_T-GNN becomes prohibitively high (994.7 seconds for NY and 685.88 seconds for CH dataset) while that of EA\_T-GNN remains reasonably low (16.5 seconds for NY and 8.7 seconds for CH dataset). This demonstrates that the exponential growth of the baseline approach makes it infeasible for processing T-GNN queries in real-time as group size increases (e.g., $n>6$). The proposed approach EA\_T-GNN demonstrates feasible execution times for processing T-GNN query even when number of users is 10 or higher.

\subsubsection{Effect of trip length}\label{ch:effect_location}
Figures~\ref{fig:varyParams}(c) and~\ref{fig:varyParams}(d) show the effect of trip length $m$ on the performance of T-GNN query processing algorithms for NY and CH datasets, respectively. For all values of $m$, EA\_T-GNN outperforms BA\_T-GNN in terms of computation time. As $m$ increases, the computation of the baseline approach increases exponentially while that of our efficient approach (EA\_T-GNN) increases linearly. For $n=10$, EA\_T-GNN approach is approx. four (4) and five (5) times faster than the baseline approach for NY and CH datasets, respectively. The exponential growth in computation time makes the baseline approach infeasible for real-time query processing for larger values of $m$, whereas EA\_T-GNN maintains a practically feasible execution time.

\subsubsection{Effect of POI density}\label{ch:effect_poi}
Figures~\ref{fig:varyParams}(e) and~\ref{fig:varyParams}f) show the effect of the POI density on the performance of T-GNN query processing algorithms for the two datasets. 
As the number of POIs (i.e, POI density) increases, the computation time of the baseline approach increases significantly. This is due to repeated computation of GTP algorithms for more number of POIs. However, the computation time of EA\_T-GNN approach does not increase much as POI density increases. The efficient pruning techniques of EA\_T-GNN enable it to compute the optimal solution by exploring only a small subset of POIs, and thus the impact of POI density is very low on its computation time. For $\rho=5$ (5\% POI density), the computation times of EA\_T-GNN are appox. five (5) and  three (3) times faster than the baseline algorithm for NY and CH datasets, respectively. 

\subsubsection{Effect of query area}\label{ch:effect_qa}
\begin{figure*}[t!]
   \begin{center}
        \begin{tabular}{cccc}
            \hspace{-6mm}

        \hspace{-4mm}
          \resizebox{40mm}{!}{
          
\begin{tikzpicture}
\begin{axis}[
 every axis label/.style={font=\Large},  
    every tick label/.style={font=\large} ,
    xlabel=$n$,
    ylabel=$Time(s)$,
    ymode = log,
    log basis y={10},
    legend pos=north west
            ]
\addplot[smooth,mark=*,color=blue, mark size=3pt] 
plot coordinates {
    (2,13.787)
    (4, 20.0772)
    (6,24.71)
    (8,67.894)
    (10,994.744)
};
\addplot[smooth,color=red,mark=x, mark size=4pt]
    plot coordinates {
    (2,	3.92866)
    (4,	7.5827)
    (6,	10.6975)
    (8,	13.7781)
    (10, 16.5733)
    };
\legend{BA\_T-GNN, EA\_T-GNN}
\end{axis}
\end{tikzpicture}
          } &
        \hspace{-4mm}
            \resizebox{40mm}{!}{
\begin{tikzpicture}
\begin{axis}[
    every axis label/.style={font=\Large},  
    every tick label/.style={font=\large} ,
    xlabel=$n$,
    ylabel=$Time(s)$,
    ymode = log,
    log basis y={10},
    legend pos=north west
            ]
\addplot[smooth,mark=*,color=blue, mark size=3pt] 
plot coordinates {
    (2,3.89735)
    (4, 5.80888)
    (6,8.65515)
    (8,35.987)
    (10,685.884)
};

\addplot[smooth,color=red,mark=x, mark size=4pt]
    plot coordinates {
    (2,	1.88426)
    (4,	3.76226)
    (6,	5.426)
    (8,	7.25534)
    (10, 8.74396)
    };
\legend{BA\_T-GNN, EA\_T-GNN}
\end{axis}
\end{tikzpicture}
}  &
   \hspace{-4mm}
          \resizebox{40mm}{!}{
\begin{tikzpicture}
\begin{axis}[
 every axis label/.style={font=\Large},  
    every tick label/.style={font=\large} ,
     xlabel=$m$,
    ylabel=$Time(s)$,
    legend pos=north west
            ]
\addplot[smooth,mark=*,color=blue, mark size=3pt] 
plot coordinates {
    (2, 6.43258)
    (4, 16.2279)
    (6,24.71)
    (8,33.2671)
    (10,77.9177)
};

\addplot[smooth,color=red,mark=x, mark size=4pt]
    plot coordinates {
    (2,	2.14547)
    (4,	6.5103)
    (6,	10.6975)
    (8, 14.2102)
    (10, 18.1659)
    };
\legend{BA\_T-GNN, EA\_T-GNN}
\end{axis}
\end{tikzpicture}
} &
\hspace{-4mm}
            \resizebox{40mm}{!}{
\begin{tikzpicture}
\begin{axis}[
 every axis label/.style={font=\Large},  
    every tick label/.style={font=\large} ,
    xlabel=$m$,
    ylabel=$Time(s)$,
    legend pos=north west
            ]
\addplot[smooth,mark=*,color=blue, mark size=3pt] 
plot coordinates {
    (2,2.4069)
    (4, 4.49893)
    (6,8.65515)
    (8,18.3186)
    (10,45.9323)
};
\addplot[smooth,color=red,mark=x, mark size=4pt]
    plot coordinates {
    (2,	1.14485)
    (4,	2.76756)
    (6,	5.426)
    (8,	7.63885)
    (10, 9.53725)
    };
\legend{BA\_T-GNN, EA\_T-GNN}
\end{axis}
\end{tikzpicture}}\\

             \hspace{-6mm}
            \hspace{10mm} \scriptsize{(a) Effect of $n$ (NY) \textsc{}\hspace{10mm}}  &  \scriptsize{(b) Effect of $n$ (CH) \textsc{}\hspace{0mm}} & \scriptsize{(c) Effect of $m$ (NY)  \textsc{}\hspace{0mm}} &  \scriptsize{(d) Effect of $m$ (CH) \textsc{}\hspace{10mm}}\\
            \hspace{-6mm}


 \hspace{-4mm}
          \resizebox{40mm}{!}{
\begin{tikzpicture}
\begin{axis}[
 every axis label/.style={font=\Large},  
    every tick label/.style={font=\large} ,
    xlabel=$\rho$,
    ylabel=$Time(s)$,
     symbolic x coords={0,0.1,0.5,1,5,10},
    xtick=data,
    legend pos=north west
            ]
\addplot[smooth,mark=*,color=blue, mark size=3pt] 
plot coordinates {
    (0.1, 12.5436)
    (0.5, 17.8798)
    (1, 24.71)
    (5, 75.819)
};

\addplot[smooth,color=red,mark=x, mark size=4pt]
    plot coordinates {
      (0.1, 9.69934)
    (0.5, 10.2532)
    (1, 10.6975)
    (5, 14.6209)
    };
\legend{BA\_T-GNN, EA\_T-GNN}
\end{axis}
\end{tikzpicture}
} &
\hspace{-4mm}
            \resizebox{40mm}{!}{
\begin{tikzpicture}
\begin{axis}[
 every axis label/.style={font=\Large},  
    every tick label/.style={font=\large} ,
    xlabel=$\rho$,
    ylabel=$Time(s)$,
    symbolic x coords={0,0.1,0.5,1,5,10},
    xtick=data,
    legend pos=north west
            ]
\addplot[smooth,mark=*,color=blue, mark size=3pt] 
plot coordinates {

     (0.1, 6.14925)
    (0.5, 7.44026)
    (1, 8.65515)
    (5, 17.7697)
};
\addplot[smooth,color=red,mark=x, mark size=4pt]
    plot coordinates {
      (0.1, 5.2706)
    (0.5, 5.3662)
    (1, 5.426)
    (5, 5.96931)
    };
\legend{BA\_T-GNN, EA\_T-GNN}
\end{axis}
\end{tikzpicture}}

       &
        \hspace{-4mm}
    

 \hspace{-4mm}
          \resizebox{40mm}{!}{
\begin{tikzpicture}
\begin{axis}[
 every axis label/.style={font=\Large},  
    every tick label/.style={font=\large} ,
    xlabel=$QA$,
    ylabel=$Time(s)$,
    symbolic x coords={25, 50, 75, 100},
    xtick=data,
    legend pos=north west
            ]
\addplot[smooth,mark=*,color=blue, mark size=3pt] 
plot coordinates {
    (25,	14.8342)
    (50,	24.71)
    (75,	33.7034)
    (100,	51.0362)
};

\addplot[smooth,color=red,mark=x, mark size=4pt]
    plot coordinates {
    (25,	9.57293)
    (50,	10.6975)
    (75,	11.9488)
    (100,	15.6499)
    };
\legend{BA\_T-GNN, EA\_T-GNN}
\end{axis}
\end{tikzpicture}
} &
\hspace{-4mm}
            \resizebox{40mm}{!}{
\begin{tikzpicture}
\begin{axis}[
 every axis label/.style={font=\Large},  
    every tick label/.style={font=\large} ,
    xlabel=$QA$,
    ylabel=$Time(s)$,
    symbolic x coords={25, 50, 75, 100},
    xtick=data,
    legend pos=north west
            ]
\addplot[smooth,mark=*,color=blue, mark size=3pt] 
plot coordinates {
    (25,	6.7311)
    (50,	8.65515)
    (75,	11.491)
    (100,	15.3833)
};
\addplot[smooth,color=red,mark=x, mark size=4pt]
    plot coordinates {
    (25,	4.71443)
    (50,	5.426)
    (75,	5.81621)
    (100,	6.51893)
    };
\legend{BA\_T-GNN, EA\_T-GNN}
\end{axis}
\end{tikzpicture}}\\

            \hspace{-6mm}
            \hspace{10mm} \scriptsize{(e) Effect of $\rho$ (NY) \textsc{}\hspace{10mm}}  &  \scriptsize{(f) Effect of $\rho$ (CH) \textsc{}\hspace{0mm}} & \scriptsize{(g) Effect of $QA$ (NY)  \textsc{}\hspace{0mm}} &  \scriptsize{(h) Effect of $QA$ (CH) \textsc{}\hspace{10mm}}\\
            \hspace{-6mm}
        \end{tabular}
       \vspace{-3mm}
        \caption{\revised{Effect of varying the group size ($n$), trip length ($m$), POI density ($\rho$), and query area ($QA$) for the New York (NY) and Chicago (CH) datasets} on the T-GNN query processing time.}
       \vspace{-7mm}
      \label{fig:varyParams}
    \end{center}
\end{figure*}
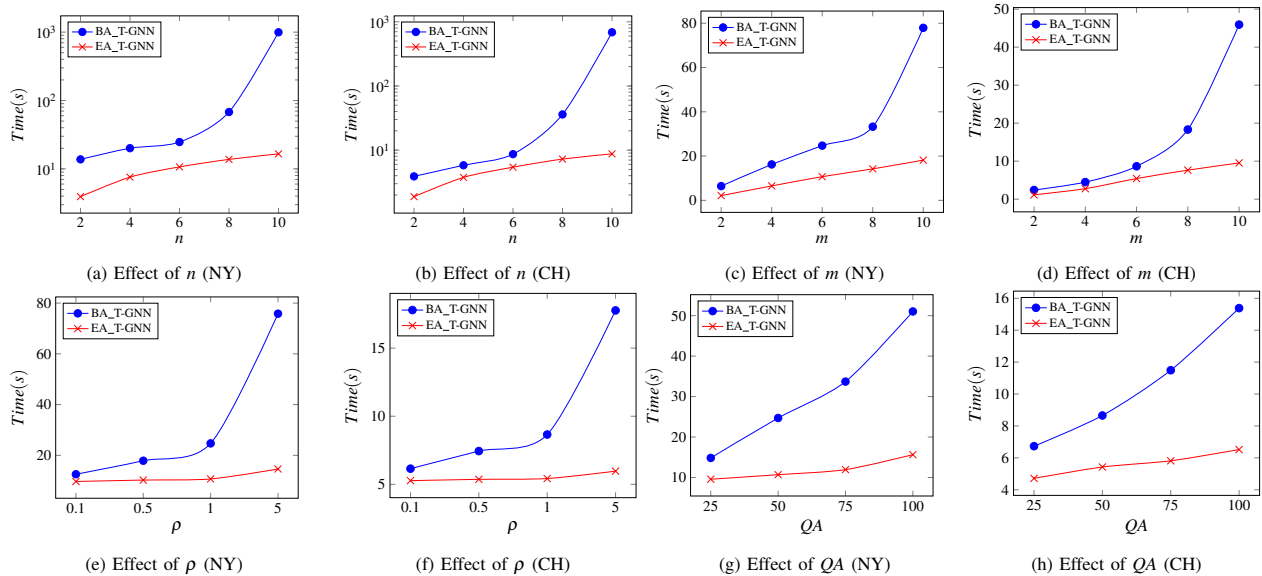

Figure~\ref{fig:varyParams}(g) and~\ref{fig:varyParams}(h) show the impact of query area on the performance of T-GNN query processing algorithms. As the query area increases (from $25\times25$ to $100\times100$), both algorithms require to explore more number of POIs to find the optimal solution resulting in an increase in processing times. However, we observe a sharp increase in processing time for the baseline approach, since it computes repeated GTP queries over the larger number of POIs. In contrast, EA\_T-GNN applies efficient pruning techniques which reduces the POI exploration to a small number of POIs. This results in a much lower computation time for EA\_T-GNN algorithm, which outperforms significantly the baseline approach.

\subsubsection{Effectiveness of pruning techniques}\label{ch:effect_pruning}
In this section, we demonstrate the effectiveness of the pruning techniques in terms of number of POIs retrieved. We conduct several experiments where all parameters are set to their default values, and we apply one or more pruning techniques (for EA\_T-GNN). 
Table~\ref{tab:pruning_poi_db} presents the experiment results. We observe that without any pruning, the performance of EA\_T-GNN drops as it explores a large number of POIs (27.9 and 8.99 seconds of processing times for NY and CH datasets, respectively). In contrast, when all pruning methods are applied, the processing times reduces significantly (10.7 and 5.4 seconds for NY and CH datasets, respectively). We observe that PT1 and PT2 are more effective than PT3 in terms of pruning efficiency. This is demonstrated by the average number of POIs retrieved in Table~\ref{tab:pruning_poi_db}, for each of the pruning techniques individually (e.g., 129.03, 128.53, and 448.10 for PT1, PT2, and PT3, respectively for NY dataset). When two pruning methods are combined, the processing time improves as the algorithm explores fewer number of POIs. Compared to the baseline method, our approach explores significantly less number of POIs even when no pruning is applied, under the default setting. The baseline approach repeatedly retrieves POIs for different combination of GTP queries resulting in substantially more POI exploration.  

\begin{small}
\begin{table}[h!]
  \begin{center}
    \caption{Effectiveness of different pruning techniques in terms of average number of POIs retrieved from database and computational time in seconds for New York (NY) and Chicago (CH) datasets}
    \label{tab:pruning_poi_db}
    \resizebox{\linewidth}{!}{
	\begin{tabular}{cccccccccc}
        \toprule
        \multicolumn{1}{c}{}  & \multicolumn{2}{c}{NY} & \phantom{ab} & \multicolumn{2}{c}{CH}\\
	\cmidrule{2-3} \cmidrule{5-6}
          & Retrieved POI & Time(s) && Retrieved POI & Time(s)  \\
        \hline
        No Pruning    & 2272 & 27.90	&& 1344	& 8.99	\\
        PT1+PT2+PT3   &\textbf{124.47}& \textbf{10.70} && \textbf{45.70}	& \textbf{5.43}	\\
        PT1    &129.03& 10.93 && 46.10	& 5.77	\\
        PT2    &128.53& 10.87 && 47.20 & 5.65	\\
        PT3    &448.10& 12.20 && 201.93	& 5.86	\\
        PT2+PT3   &126.17&	11.06 && 47.20 & 5.61\\
        PT1+PT3   &126.60&	11.14 && 46.10 & 5.71\\
        PT1+PT2   &126.83&	10.82 && 45.70 & 5.72\\
        Baseline      &321001.40 & 24.71 && 107756.50 & 8.66\\
       \bottomrule
    \end{tabular}}
  \end{center}
\end{table}
\end{small}

\subsubsection{Scalability of EA\_T-GNN}\label{ch:scalability}
In this section, we evaluate the scalability of the EA\_T-GNN approach with respect to the group size. Specifically, we set the group size ($n$) to larger values (ranging from 5 to 30) than those used in previous experiments and examine the impact on the performance of T-GNN query processing. The results are summarized in Table~\ref{tab:scalability} for different POI densities. We observe that for the highest setting ($n = 30$ and $\rho=5\%$), the computation times of EA\_T-GNN are 53.72 and 26.41 seconds for the New York and Chicago datasets, respectively. The results demonstrate that EA\_T-GNN is capable of processing T-GNN queries in near real-time—even for large groups of up to 30 users. In practice, group sizes are typically smaller, supporting the scalability of the proposed approach for real-time applications.

\begin{small}
\begin{table}[h!]
  \begin{center}
    \caption{Scalability of EA\_T-GNN shown in terms of group size. Computational time in seconds is reported for varying group size and POI density}
    \label{tab:scalability}
    \resizebox{\linewidth}{!}{
	\begin{tabular}{cccccccccc}
        \toprule
        \multicolumn{1}{c}{}  & \multicolumn{4}{c}{NY} & \phantom{ab} & \multicolumn{4}{c}{CH}\\
	\cmidrule{2-5} \cmidrule{7-10}
        $n$& $0.1\%$ & $0.5\%$ & $1\%$ & $5\%$ && $0.1\%$ & $0.5\%$ & $1\%$ & $5\%$ \\
        \hline
        5 & 7.98& 8.37	& 8.76	& 12.39	&& 4.41	& 4.67	& 4.76	& 5.61\\
        10 &15.92& 16.38& 16.57	& 21.20	&& 8.88	& 8.93	& 8.74	& 9.62\\
        15 &23.59& 24.25& 24.83	& 30.87	&& 13.08& 13.18	& 13.39	& 14.42\\
        20 &30.23& 31.28& 31.83 & 39.31	&& 17.06& 16.96 & 17.40 & 18.25\\
        25 &36.98& 37.86& 38.76	& 48.09	&& 20.62& 20.79	& 20.74	& 21.60\\
        30 &45.33&	46.41&	46.83 & 53.72 && 24.59&	24.72 &	25.07 & 26.41\\





       \bottomrule
    \end{tabular}}
  \end{center}
\end{table}
\end{small}

\section{Conclusions}{\label{conclusion}}
In this paper, we introduce a novel query called Trip-based Group Nearest Neighbor (T-GNN) query. Given the predefined trips of users in a group, the T-GNN query returns the optimal meetup POI for the group along with the detour locations. The additional distance traveled by the user due to the detour from their original trip is termed as overhead distance. We present an efficient approach to process T-GNN queries in real-time that find the optimal meetup POI minimizing the total trip overhead distance for the group. Our approach incrementally retrieves POIs from the database according to distance from a centroid, continuously updates the current best solution, and prunes the search area based on the computed solution. We develop efficient pruning techniques that substantially prunes POIs from the candidate set, improving query processing efficiency of the proposed approach significantly. We conducted extensive experiments to validate the performance of our approach on two real-world road network datasets. Our approach outperforms a baseline algorithm with reduced computational time, fewer POI exploration, and enhanced scalability. Future works include considering user preferences in T-GNN query, where users rank locations in their trip denoting their preference of detour locations. This assists the group to generate a better meetup POI minimizing overhead distance and optimizing detour locations according to user preferences.

\bibliographystyle{IEEEtran}
\bibliography{IEEEabrv,references}

\begin{IEEEbiography}[{\includegraphics[width=1in,height=1.25in,clip,keepaspectratio]{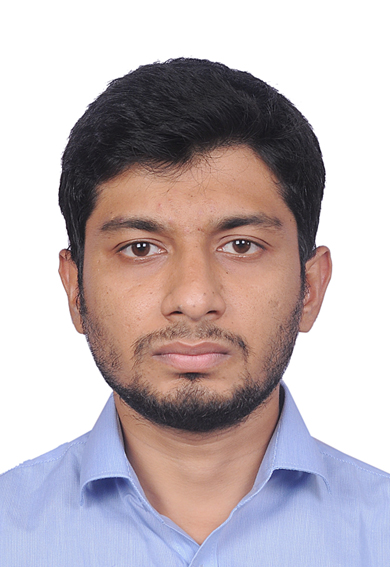}}]{Shahiduz Zaman}
Shahiduz Zaman is working at DESCO as an Assistant Engineer. Shahiduz Zaman completed his B.Sc. and M.Sc. degree from Bangladesh University of Engineering and Technology (BUET). His research interest falls in the area of spatial database, human-computer interaction, and machine learning. 
\end{IEEEbiography}

\begin{IEEEbiography}[{\includegraphics[width=1in,height=1.25in,clip,keepaspectratio]{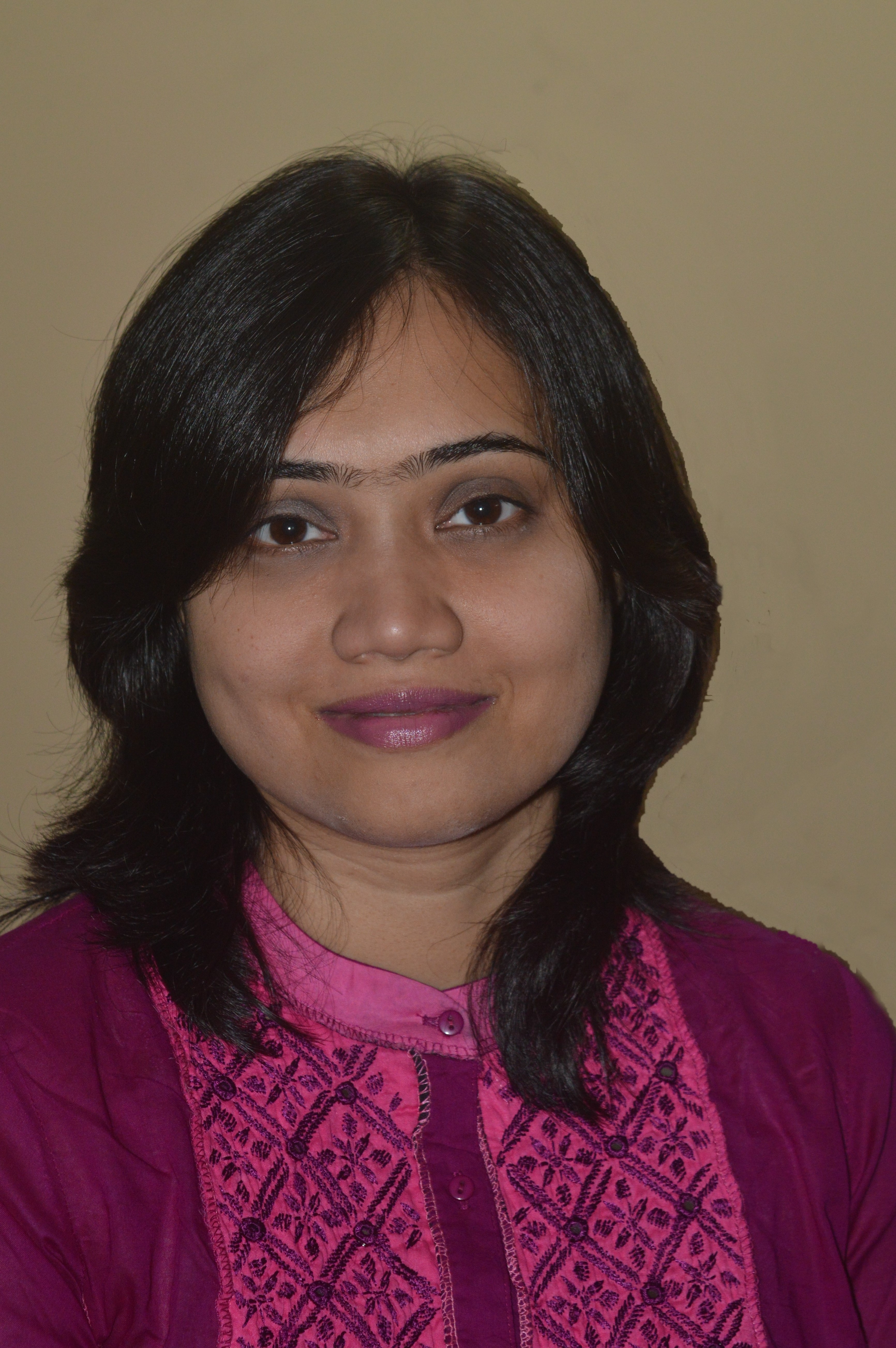}}]{Tanzima Hashem}
Dr. Tanzima Hashem is a professor at the Department of Computer Science and Engineering of Bangladesh University of Engineering and Technology (BUET). She received her Ph.D. degree from the University of Melbourne, Australia in 2012. Her research interest falls in the area of ubiquitous computing, spatial databases and privacy. In 2017, she received the prestigious OWSD-Elsevier Foundation Award for Early-Career Women Scientists in the Developing World in Engineering Sciences.
\end{IEEEbiography}

\begin{IEEEbiography}[{\includegraphics[width=1in,height=1.25in,clip,keepaspectratio]{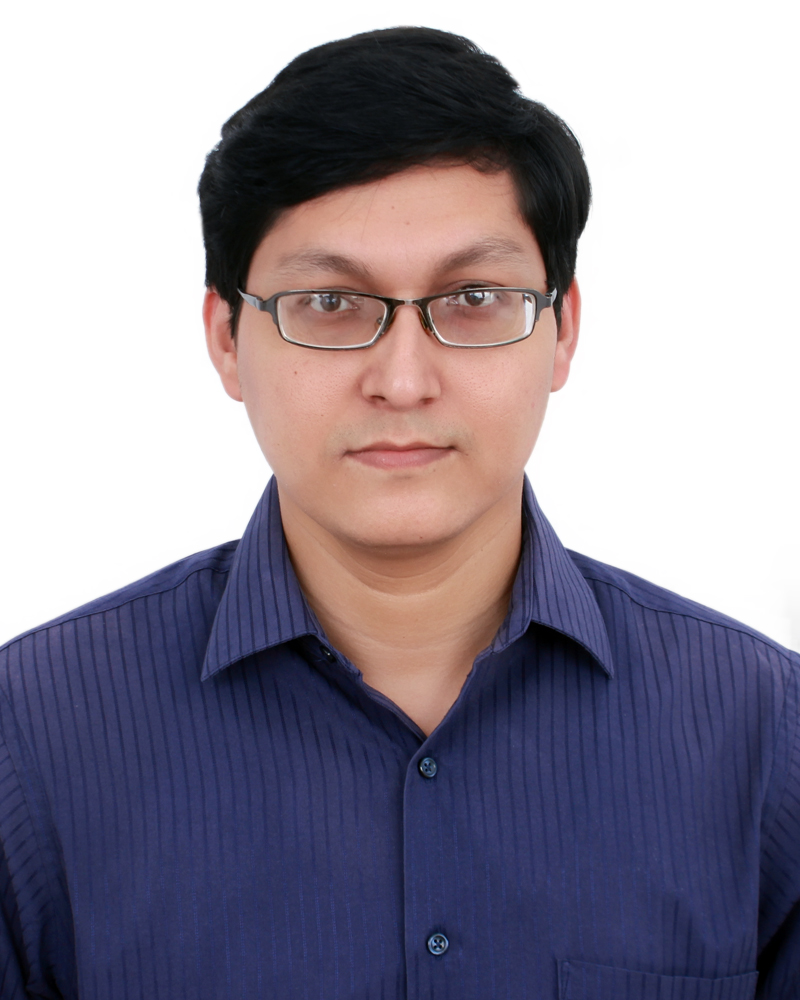}}]{Sukarna Barua}
Sukarna Barua received the B.Sc. Engg. and M.Sc. Engg. degrees in computer science and engineering from the Bangladesh University of Engineering and Technology (BUET), Dhaka, Bangladesh, in 2009 and 2011 respectively. He is currently an associate professor in the Department of Computer Science and Engineering, BUET. His research interests include spatial query processing, machine learning, and deep learning.
\end{IEEEbiography}

\vfill

\end{document}